\documentclass[preprint,10pt]{elsarticle}

\newcounter{bla}
\newenvironment{refnummer}{%
\list{[\arabic{bla}]}%
{\usecounter{bla}%
 \setlength{\itemindent}{0pt}%
 \setlength{\topsep}{0pt}%
 \setlength{\itemsep}{0pt}%
 \setlength{\labelsep}{2pt}%
 \setlength{\listparindent}{0pt}%
 \settowidth{\labelwidth}{[9]}%
 \setlength{\leftmargin}{\labelwidth}%
 \addtolength{\leftmargin}{\labelsep}%
 \setlength{\rightmargin}{0pt}}}
 {\endlist}

\journal{Computer Physics Communications}

\usepackage{graphicx}% Include figure files
\usepackage{dcolumn}% Align table columns on decimal point
\usepackage{bm}% bold math
\usepackage{booktabs}
\usepackage{enumerate}
\usepackage{pgf,tikz}
\usetikzlibrary{arrows}
\usepackage{amssymb}

\usepackage{color}
\usepackage{siunitx}
\usepackage{multirow}
\usepackage{threeparttable}
\usepackage[normalem]{ulem}
\usepackage{amsmath}
\usepackage{booktabs}   % table rules
\usepackage{makecell}   % table header line breaks
\usepackage{rotating}
\usepackage{url}
\usepackage{listings}
\usepackage{float}
\usepackage[hidelinks]{hyperref}% clickable TOC and refs

\begin{document}

\begin{frontmatter}


%% and the form \ead[nolinkurl] for the home page:
%% \ead[nolinkurl]{home page}

\title{A CSFG-based neural network basis-selection method for large-scale RCI calculations within {\sc Graspg}}

%\title{A CSFG-based neural network configuration-selection method for large-scale RCI calculations}
%\title{Neural-Network-Driven Configuration Space Reduction Method based on CSFG for Large-Scale RCI Calculations}


\author[a]{Chaofan Shi}
\author[a]{Shaowei Tian}
\author[a]{Ran Si\corref{author}}
\author[b]{Gediminas Gaigalas}
\author[c]{Per J\"onsson}
\author[a]{Chongyang Chen\corref{author}}

\cortext[author] {Corresponding author.\\\textit{E-mail addresses:} rsi@fudan.edu.cn, chychen@fudan.edu.cn}

\address[a]{Institute of Modern Physics, Fudan University, Shanghai 200433, China}
\address[b]{Institute of Theoretical Physics and Astronomy, Vilnius University, Saulėtekio av. 3, LT-10222 Vilnius, Lithuania}
\address[c]{Department of Materials Science and Applied Mathematics, Malmö University, SE-20506 Malmö, Sweden}

% Additional affiliations can be added here if needed.

\date{\today}% It is always \today, today,

\begin{abstract}
We present a neural network (NN) basis-selection method for large-scale relativistic configuration interaction (RCI) calculations in {\sc Graspg}. The method employs configuration state function generators (CSFGs), each of which generates a set of configuration state functions (CSFs) with the same spin-angular couplings, as the basic selection units for the NN.
A constant-orbital feature-elimination strategy removes feature channels whose values remain unchanged across the CSFG pool. The CSFG representation reduces the number of learning units processed by the NN by more than one order of magnitude, while constant-orbital feature elimination further reduces the dimensionality of the NN input. Combined with the high-performance {\sc Graspg} framework, the method improves the efficiency of both NN selection and subsequent RCI calculations, maintaining a balance between accuracy and computational cost.
In a moderate Ni\(^{12+}\) benchmark, where the corresponding full-space RCI calculation is still feasible, the CSFs generated by the retained CSFG sets reproduce the full-space RCI results at the few \(\mathrm{cm}^{-1}\) level for the target states. For the representative \(0^{+}\) block, the complete workflow reduces the wall time by 75.6\%, and the peak memory required by a single RCI calculation is reduced by a factor of 10.1. In a larger-scale calculation with a full CSF expansion containing \(1.27\times10^{9}\) CSFs, the method retains only 1.1\%--1.9\% of the full-space CSFs and yields energy levels in good agreement with experimental data and other resource-intensive theoretical calculations.

% A submitted program is expected to satisfy the following criteria: it must be of benefit to other physicists, or be an exemplar of good programming practice, or illustrate new or novel programming techniques which are of importance to computational physics community; it should be implemented in a language and executable on hardware that is widely available and well documented; it should meet accepted standards for scientific programming; it should be adequately documented and, where appropriate, supplied with a separate User Manual, which together with the manuscript should make clear the structure, functionality, installation, and operation of the program.

% Your manuscript and figure sources should be submitted through Editorial Manager (EM) by using the online submission tool at \\
% https://www.editorialmanager.com/comphy/.

% In addition to the manuscript you must supply: the program source code; a README file giving the names and a brief description of the files/directory structure that make up the package and clear instructions on the installation and execution of the program; sample input and output data for at least one comprehensive test run; and, where appropriate, a user manual.

% A compressed archive program file or files, containing these items, should be uploaded at the "Attach Files" stage of the EM submission.

% For files larger than 1Gb, if difficulties are encountered during upload the author should contact the Technical Editor at cpc.mendeley@gmail.com.
% \\

\vspace{2mm}
\noindent \textbf{PROGRAM SUMMARY}

\begin{small}
\begin{sloppypar}
\noindent
{\em Program Title:} CSFG-based NN basis-selection toolkit for {\sc Graspg} \\
{\em CPC Library link to program files:} (to be added by Technical Editor) \\
{\em Developer's repository link:} Not available; the source code is supplied as supplementary material \\
{\em Licensing provisions:} MIT License\\
{\em Programming language:} Fortran 90/95, Python 3, and Unix shell scripts \\
{\em Supplementary material:} {\sc Graspg} patch file, Python toolkit, automated workflow scripts, README file, and Ni\(^{12+}\) example input files \\
{\em External routines/libraries:} {\sc Grasp}2018 and {\sc Graspg}; MPI; Python packages NumPy, TensorFlow, and psutil \\
{\em Nature of problem:}
Accurate atomic state functions (ASFs) for complex systems may require expansions containing hundreds of millions to billions of configuration state functions (CSFs), making direct full-space relativistic configuration interaction (RCI) calculations impractical in memory, storage, and wall time. Efficient selection of CSFs is therefore needed to identify and retain the CSFs carrying the dominant contributions to the target ASFs. \\
{\em Solution method:}
The package combines the CSFG-based {\sc Graspg} framework [1,2] with an NN-supported selection strategy [3]. The module \nolinkurl{nn_load_csfg_mpi} produces compact CSFG features based on a constant-orbital feature
elimination strategy, while \nolinkurl{rmixcsfg_gm} computes CSFG weights from the RCI mixing coefficients and determines adaptive cumulative-weight cutoffs that define the NN importance labels. Using these features and labels, the Python workflow trains the CNN on the current CSFG set, predicts important CSFGs in the remaining pool, and iteratively updates the working CSFG set until the target-state energies converge. \\
{\em Additional comments including restrictions and unusual features:}
The workflow uses the {\sc Graspg} file format and requires MPI-enabled {\sc Graspg} executables. {\sc Graspg}-side preprocessing and RCI calculations are CPU/MPI based, while the NN stage can run on CPU or GPU depending on the TensorFlow environment.\\

\noindent \textbf{References:}
\begin{refnummer}
\item Y. T. Li, K. Wang, R. Si, M. Godefroid, G. Gaigalas, C. Y. Chen, and P. J{\"o}nsson,
Comput. Phys. Commun. 283 (2023) 108562.
\item R. Si, Y. Li, K. Wang, C. Chen, G. Gaigalas, M. Godefroid, and P. J{\"o}nsson,
Comput. Phys. Commun. 312 (2025) 109604.
\item P. Bilous, C. Cheung, and M. S. Safronova,
Comput. Phys. Commun. 315 (2025) 109731.
\end{refnummer}

\end{sloppypar}
\end{small}
\end{abstract}
\end{frontmatter}

% \clearpage
% \tableofcontents
% \clearpage

\clearpage


%% First section: Introduction
\section{\label{introduction}Introduction}

Atomic spectroscopy is an essential tool in modern natural science, with applications in astrophysics, plasma physics, atomic and nuclear physics, precision metrology, and tests of fundamental interactions~\cite{Bailey2015,Ullmann2017,Sanner2019}. In recent years, high-precision atomic data have also become central to atomic clocks, nuclear clocks, and searches for new physics beyond the Standard Model~\cite{Godun2014,King2022,Zhang2024NuclearClock}. Meeting these demands for high-precision atomic data requires an accurate description of electron correlation at a computationally tractable cost, which remains a central challenge in many-electron atomic calculations.

The GRASP family of programs is one of the most widely used toolkits for relativistic atomic-structure calculations of many-electron systems~\cite{Jonsson2023}. Based on the multiconfiguration Dirac-Hartree-Fock (MCDHF) and relativistic configuration interaction (RCI) methods, GRASP expands atomic state functions (ASFs) as linear combinations of configuration state functions (CSFs) to describe electron correlation~\cite{FroeseFischer2016}. {\sc Graspg} extends this framework by introducing configuration state function generators (CSFGs), which exploit repeated spin-angular couplings to reduce spin-angular integrations, disk storage, memory requirements, and execution time in large-scale MCDHF and RCI calculations~\cite{Li2023,Si2025}. Nevertheless, the size of the CSF expansion grows rapidly with the number of correlated electrons and the size of the active orbital set. For complex systems, the number of CSFs can reach hundreds of millions to billions as the active orbital set is enlarged, making direct full-space RCI calculations impractical even within the CSFG framework.

Since only a small fraction of CSFs or CSFGs contribute significantly to the target ASFs, some selected-RCI and condensation strategies have been developed to remove weakly contributing components. For example, the CSFG-based condensation method in {\sc Graspg} performs a small-scale RCI calculation, identifies important CSFGs by accumulating the squared mixing coefficients of the generated CSFs, and then extends the selected CSFGs to a larger orbital set by increasing the principal quantum numbers~\cite{Li2023}. Zhang \emph{et al.} have developed a variational Monte Carlo RCI approach within the MCDHF framework, which iteratively selects important CSFs via Monte Carlo sampling and a rejection-acceptance mechanism based on energy evaluation~\cite{Zhang2024}.

Beyond these a priori or stochastic selection strategies, neural networks (NNs) provide a data-driven route to selecting important components from large CSF sets. Machine learning has already proven to be effective for configuration selection in molecular calculations~\cite{Coe2018,Jeong2021}. Recently, Bilous \emph{et al.} introduced deep learning methods into atomic calculations~\cite{Bilous2023}. Working with the {\sc Grasp}2018 program package~\cite{FroeseFischer2019}, they employed a convolutional neural network (CNN) to learn CSF features from smaller calculations, predict important CSFs, and iteratively enlarge the selected set until the results approached those of the full-space calculation. Subsequently, Bilous \emph{et al.} extended this method to the pCI program~\cite{Bilous2024}, introduced further improvements, and developed a Python package for NN-supported large-scale CI calculations using pCI and other high-performance atomic codes~\cite{Bilous2025}. These previous works demonstrated the potential of NN-based basis selection for atomic calculations and laid the methodological foundation for the present work. 

Previous CSFG-based developments have shown that the CSFG framework can greatly reduce the memory, disk, and CPU-time requirements of large-scale MCDHF and RCI calculations compared with conventional CSF-based implementations~\cite{Li2023,Si2025,Li2025}. This motivates moving the NN selection unit from individual CSFs to CSFGs. Since each CSFG represents a set of CSFs linked by spin-angular-preserving de-excitations, CSFG-level selection significantly reduces the number of selection units. In this work, we integrate NN-supported basis selection with {\sc Graspg} and develop a CSFG-based NN method to obtain a reduced CSFG set for large-scale RCI calculations. Consequently, the NN stage handles far fewer learning units, and the subsequent RCI calculation is performed in a compact CSF space generated by the selected CSFGs within the efficient {\sc Graspg} framework. The algorithm further adopts an adaptive cumulative truncation strategy to dynamically determine the importance cutoff, avoiding empirical threshold tuning and enhancing generality. Random exploration is introduced during the iterations to keep the training set exploratory and to reduce the risk of convergence to an incomplete CSFG set. The workflow is implemented through two {\sc Graspg} extension modules and a Python toolkit, and the source code together with example input files are provided as open-source supplementary material to ensure reproducibility.

The remainder of this paper is organized as follows. Section~\ref{Theory and Methods} briefly introduces the MCDHF and RCI methods, the CSFG concept, and existing resource-reduction strategies for large-scale RCI calculations in the {\sc Graspg} package. Section~\ref{sec:nn-graspg} describes in detail the proposed CSFG-based NN selection method. Section~\ref{Structure and installation} presents the structure of this NN selection package and its installation procedure. Section~\ref{sec:running-graspg} demonstrates the use of the method with Ni\(^{12+}\) as an example and analyzes its accuracy and computational efficiency. Section~\ref{Summary} summarizes the present work. All calculations were performed on a Rocky Linux 10.0 CPU server equipped with two AMD EPYC 7763 64-core processors (128 CPU cores in total) and 2.0 TB of memory. 

%% second section: theory
\section{\label{Theory and Methods}Theory and Methods}

\subsection{\label{MCDHF and RCI}MCDHF and RCI}

In the relativistic multiconfiguration method implemented in GRASP, an atomic state function (ASF) is expanded as a linear combination of CSFs with the same parity \(P\), total angular momentum \(J\), and projection \(M\):
\begin{equation}
\Psi_k(\gamma_k PJM) = \sum_{r=1}^{N_C} c_{rk} \Phi_r(\gamma_r PJM),
\label{eq:asf_expansion}
\end{equation}
where \(\gamma\) denotes additional quantum numbers, \(N_C\) is the number of CSFs, and \(c_{rk}\) are the mixing coefficients. The CSFs are constructed from one-electron Dirac orbitals~\cite{FroeseFischer2016}. In MCDHF calculations, the Dirac-Coulomb (DC) Hamiltonian \({H}_{\mathrm{DC}}\) is considered. Substituting the expansion into the eigenvalue equation,
\begin{equation}
{H}_{\mathrm{DC}} \Psi_k = E_k \Psi_k,
\label{eq:dc_eigenvalue}
\end{equation}
leads to a Hamiltonian matrix eigenvalue problem, whose diagonalization gives the energies \(E_k\) and mixing coefficients \(c_{rk}\). In the MCDHF method, the radial orbitals and mixing coefficients are optimized self-consistently by applying the variational principle to a weighted energy functional of the targeted states.

In the RCI calculation, the radial orbitals obtained from the preceding MCDHF calculation are kept fixed. The task is then reduced to solving the eigenvalue problem for a prescribed CSF expansion. This stage allows the Hamiltonian to be extended, for example, by including Breit and QED corrections, while the CSF expansion can be enlarged by adding further excitations and multi-reference (MR) configurations to improve the treatment of electron correlation.

\subsection{\label{CSFG}CSFG}
The concept of CSFG was introduced by Li \emph{et al.}~\cite{Li2023}, based on the fact that spin-angular integrations are independent of the principal quantum number~\cite{Gaigalas1997}. The CSF space is partitioned into a labeling space and a correlation space. CSFs in the labeling space are generated from the MR configurations by applying a broad range of excitations (SDTQ etc.) to a finite set of highly occupied labeling-ordered orbitals (e.g., $1s, 2s, 2p-, 2p, 3s, \ldots$), capturing dominant correlation effects arising from near-degeneracy and long-range rearrangement. The CSFs in the correlation space are generated from the MR by SD excitations to symmetry-ordered (SO) orbitals (e.g., $s, p-, p, d-, \ldots$). They primarily describe dynamic correlation effects and are typically far more numerous than the labeling-space CSFs.

A CSFG is defined as a CSF generator in which the SO orbitals are assigned their highest principal quantum numbers. From a single CSFG, a group of CSFs is generated by de-exciting the SO orbital(s) to lower principal quantum numbers, while keeping all spin-angular coupling quantum numbers unchanged. Once the spin-angular integration between two CSFGs has been evaluated, the same spin-angular coefficients can be reused for all CSF pairs generated from these two CSFGs. This avoids repeated spin-angular integrations, thereby accelerating Hamiltonian construction and reducing the amount of data stored in memory and on disk. Moreover, the number of CSFGs is independent of the highest principal quantum numbers of the orbitals in the SO set, making the CSFG representation especially attractive for large active sets containing many correlation orbitals of the same symmetry.

Using the CSFG-based {\sc Graspg} package, MCDHF calculations have achieved reductions in execution time and disk file size by factors of 37 and 98, respectively, compared with conventional CSF-based {\sc Grasp}2018 calculations. For RCI calculations, speedups by factors exceeding 200 have been reported~\cite{Si2025,Li2023}.

\subsection{\label{For Large-Scale Calculations}Existing resource-reduction strategies in {\sc Graspg}}

In principle, an exact ASF would require a complete CSF expansion. In practice, high-precision calculations approach this limit by increasing the active orbital set and upgrading the virtual excitation model. This improves the description of correlation but rapidly increases storage, memory usage, and execution time. Several approximations have therefore been developed in {\sc Graspg} to reduce computational cost while preserving the dominant correlation effects.

\subsubsection{\label{Zero-First-Order Method}Zero-first-order method}

The zero-first-order (ZF) method~\cite{Jonsson2014,Gustafsson2017} divides the CSF expansion into two subspaces. The \(P\) space (zero-order space) contains the CSFs that contribute most strongly to the wave function, such as MR configurations and important CSFs generated by excitations from reference configurations. The \(Q\) space (first-order space) contains a much larger number of weakly contributing CSFs.

When constructing the Hamiltonian matrix, the \(P\)-\(P\) block, the \(P\)-\(Q\) coupling blocks, and only the diagonal elements of the \(Q\)-\(Q\) block are kept. This approximation greatly reduces the number of matrix elements and therefore the memory usage and execution time. The resulting energies are similar to those obtained by applying second-order Brillouin-Wigner perturbative corrections to the zero-order matrix \(H^{(PP)}\).

By employing the CSFG concept, Li \emph{et al.}~\cite{Li2025} introduced a blockwise ZF method in which \(H^{(QQ)}\) is treated as a block-diagonal matrix. Each block contains the interactions among all CSFs generated by a single CSFG, rather than only the diagonal elements within \(H^{(QQ)}\). For W\(^{37+}\), the blockwise ZF calculation with valence-valence (VV) correlation placed in the zero-order space agrees with the corresponding full-interaction calculation within the same CSF expansion to within 0.02\%, while reducing the number of non-zero matrix elements from approximately 5.3 billion to about 1.3 billion~\cite{Li2025}.

However, ZF and blockwise ZF reduce the cost of treating the Hamiltonian matrix rather than the size of the underlying CSF expansion itself. If the zero-order space is not sufficiently small, even a ZF calculation can remain infeasible; if it is made too small, important correlation effects may be lost. Therefore, a complementary strategy that reduces the CSF expansion itself is still needed.

\subsubsection{\label{Condensation}A priori condensation}

Based on the CSFG concept, an a priori condensation method has also been proposed within {\sc Graspg} to remove CSFGs with small contributions to the target ASFs. The condensation procedure proceeds as follows. First, a small-scale RCI calculation is performed on a limited orbital set to obtain estimates of the CSF mixing coefficients.

For \(N_{\mathrm{state}}\) target states belonging to the same \(J^{\pi}\) block, the total normalized weight carried by the labeling-space CSFs is defined as
\begin{equation}
W_{\mathrm{label}}
=
\frac{1}{N_{\mathrm{state}}}
\sum_{k=1}^{N_{\mathrm{state}}}
\sum_{r\in\mathcal{I}_{\mathrm{label}}}
\left|c_{rk}\right|^2 ,
\label{eq:label_weight}
\end{equation}
where \(\mathcal{I}_{\mathrm{label}}\) denotes the set of all labeling-space CSFs, and \(c_{rk}\) is the mixing coefficient of CSF \(r\) in target state \(k\). For each correlation-space CSFG \(g\), its normalized weight is defined as
\begin{equation}
W_g
=
\frac{1}{N_{\mathrm{state}}}
\sum_{k=1}^{N_{\mathrm{state}}}
\sum_{r\in\mathcal{I}_g}
\left|c_{rk}\right|^2 ,
\label{eq:csfg_weight}
\end{equation}
where \(\mathcal{I}_g\) contains all CSFs generated from CSFG \(g\). Since every target ASF is normalized and the labeling- and correlation-space CSFs together form the whole CSF expansion, these weights satisfy
\begin{equation}
W_{\mathrm{label}}
+
\sum_{g\in G_{\mathrm{corr}}} W_g
=1 ,
\label{eq:weight_normalization}
\end{equation}
where \(G_{\mathrm{corr}}\) denotes the set of correlation-space CSFGs.

The CSFGs are then sorted in descending order of \(W_g\) and accumulated until a predefined cumulative fraction is reached. The CSFGs contributing to this cumulative sum are classified as important, while the rest are discarded. Because CSFGs are defined by spin-angular patterns rather than by a fixed maximum principal quantum number, the resulting CSFG list can be extended to larger orbital sets by increasing the principal quantum numbers of the SO correlation orbitals. This a priori condensation significantly reduces the number of CSFs that need to be handled in later RCI calculations.

This strategy is powerful when the preliminary small-scale calculation is affordable and representative. Its limitation is that the condensation threshold has to balance cost and accuracy: retaining too many CSFGs may still leave an infeasible large-scale calculation, while retaining too few may remove important correlation effects. More adaptive approaches are therefore desirable when the relevant large CSF space is already beyond the reach of a reliable preliminary condensation calculation.

%% third section: NN-supported RCI algorithm based on CSFG
\section{\label{sec:nn-graspg}CSFG-based NN selection method}
When the CSF space exceeds the practical limits of direct full-space RCI calculations, the central task is to select a compact subset that preserves the important correlation effects. Machine learning, and neural networks (NNs) in particular, provide a way to generate this subset from the results of smaller RCI calculations. Bilous \emph{et al.}~\cite{Bilous2023} implemented an NN-supported algorithm based on {\sc Grasp}2018. Their approach starts from a small subset of the full CSF set, performs an RCI calculation on the current working set, trains the NN to distinguish important CSFs from unimportant ones, and iteratively updates the working set until the target energies converge.

In this work, the selection unit is changed from CSF to CSFG. This choice matches the internal structure of {\sc Graspg}. We adopt the CNN architecture of the earlier NN-supported RCI work but redesign the data flow, labeling strategy, and iteration management on the basis of the CSFG representation. {\sc Graspg} is extended with \nolinkurl{nn_load_csfg_mpi} and \nolinkurl{rmixcsfg_gm}, and the Python front-end is provided through \nolinkurl{Feature_Transform.py}, \nolinkurl{Iteration_manage.py}, and \nolinkurl{nn_graspg_toolkit.py}. Together, these components form a CSFG-based NN selection method deeply integrated with {\sc Graspg} and suitable for RCI calculations containing hundreds of millions to billions of CSFs.

\subsection{\label{Features of CSFGs}CSFG features}
In this algorithm, CSFG features serve as input to the NN. As discussed in Section~\ref{CSFG}, a CSFG can be regarded as the representative CSF with the highest principal quantum numbers among CSFs of a specified type. In the present implementation, each CSFG is described by three feature streams: orbital occupation numbers, subshell angular quantum numbers, and intermediate coupled angular momenta. These streams form the input channels of a one-dimensional (1D) convolutional layer. It should be noted that the present three-stream input cannot uniquely represent CSFGs involving multiply occupied high-angular-momentum subshells that require seniority, quasispin, or other additional quantum numbers~\cite{Jonsson2023}.

Here we further optimize the feature preprocessing in two ways. First, Bilous \emph{et al.} showed that a CNN can automatically suppress the background formed by fully occupied and vacant orbitals in CSFs, indicating that such orbitals are weakly correlated with CSF importance (see Ref.~\cite{Bilous2023} and its Supplemental Material). In large-scale calculations, the practical issue is that such redundant channels still consume disk space, memory, and prediction time even if the CNN can eventually learn to ignore them. Moreover, in the CSFG representation, more orbitals have occupation patterns that remain unchanged across the full CSFG set. We therefore remove the feature channels associated with these constant orbitals. We refer to this preprocessing step as constant-orbital feature elimination. It reduces the input dimensionality and removes redundant background information before NN training and prediction.

Second, in the original algorithm~\cite{Bilous2023}, CSF features are read from CSF text files, normalized, and transformed to NumPy arrays in Python, which is extremely time-consuming when handling tens of millions of CSFs. To address this, we developed a parallel {\sc Graspg}-side loading module named \nolinkurl{nn_load_csfg_mpi}, which reads CSFG files, removes constant-orbital feature channels, extracts compact CSFG features, and stores them in the binary file \nolinkurl{compact_listb.dat}. The Python module \nolinkurl{Feature_Transform.py} then reads this binary file and writes the CSFG-pool feature array used by the NN. It should be noted that the CSFG features are not normalized in the present workflow; instead, the original integer-valued descriptors are used directly and stored with one byte per descriptor (\texttt{BYTE}-level storage). The same compact integer representation is preserved in the Python feature array used by the NN, substantially reducing the memory footprint and disk usage compared with floating-point feature arrays.

As an example, the even \(J=1\) block of Ni\(^{12+}\) discussed in Section~\ref{Application to realistic problem} contains 437\,285\,546 generated CSFs and 115 relativistic orbitals in the CSF-based representation.  Table~\ref{tab:csfg_csf_list} gives a representative CSFG-list excerpt in the {\sc Graspg} format~\cite{Si2025}. The first orbitals under the peel-subshells heading are in labeling order, while the remaining orbitals are arranged in symmetry order and form the correlation-orbital part. It also shows a sample CSFG with the highest-\(n\) pair \(10p11p\), which represents 21 generated CSFs from all \(npn'p\) pairs with \(5\le n<n'\le 11\). This one-to-many correspondence provides an important motivation for using CSFGs, rather than individual CSFs, as the learning units in the NN selection. In this representation, the lower-\(n\) correlation orbitals generated by these de-excitations are not explicitly occupied in the representative CSFG, so many corresponding orbital-feature channels remain vacant or constant across the CSFG pool and can be removed by constant-orbital feature elimination.

\begin{table}[H]
\centering
\caption{Representative CSFG-list excerpt for the even \(J=1\) block of Ni\(^{12+}\), showing the labeling-space CSFs, correlation-space CSFGs, and the sample CSFG discussed in the text.
} 
\label{tab:csfg_csf_list}
\begin{minipage}{0.98\textwidth}
\begin{lstlisting}[basicstyle=\ttfamily\scriptsize,columns=fullflexible,keepspaces=true]
Core subshells:

Peel subshells:
  1s   2s   2p-  2p   3s   3p-  3p   3d-  3d   4s   4p-  4p
  5s   6s   ...  11s  5p-  ...  11p- 5p   ...  11p  ...
  ...
  9l-  10l- 11l- 9l   10l  11l

CSF(s):
  # start CSFs in labeling space
  1s ( 2)  2s ( 2)  2p-( 2)  2p ( 4)  3s ( 2)  3p-( 1)  3p ( 3)
                                                   1/2      3/2
                                                               1+
  ....
  2s ( 2)  2p-( 2)  2p ( 4)  3p ( 4)  4p-( 1)  4p ( 3)
                                             1/2      3/2
                                                         1+
  # end CSFs in labeling space

  # start CSFGs in correlation space
  1s ( 2)  2s ( 2)  2p-( 2)  2p ( 4)  3s ( 2)  3p-( 2)  3p ( 1)  11p-( 1)
                                                            3/2      1/2
                                                                        1+
  ....
  # example CSFG
  1s ( 2)  2s ( 2)  2p-( 2)  2p ( 4)  3s ( 2)  3p-( 2) 10p ( 1) 11p ( 1)
                                                            3/2      3/2
                                                                        1+
  # end example CSFG
  ....
  2s ( 2)  2p-( 2)  2p ( 4)  3p ( 4)  4p ( 2)  11l ( 2)
                                             2        2
                                                         1+
  # end CSFGs in correlation space
\end{lstlisting}
\end{minipage}
\end{table}

After switching to the CSFG-level representation and applying constant-orbital feature elimination, the basis size is reduced to 15\,056\,337 CSFGs, and only 46 relativistic orbitals are retained in the feature representation. As summarized in Table~\ref{tab:feature_comparison}, the number of learning units is reduced by a factor of 29.0, while the orbital-feature dimension is reduced by a factor of 2.5. Since each non-primary CSF or CSFG is represented by three descriptor streams, the total feature count is given by \((N_{\mathrm{basis}}-N_{\mathrm{primary}}) N_{\mathrm{orb}}\times 3\), giving an overall feature-count reduction by a factor of 72.6. The corresponding NumPy array size follows from this feature count and the data type used for storage. Therefore, the change from floating-point CSF feature arrays to compact \texttt{BYTE}/\texttt{int8} CSFG feature arrays reduces the array size from about 603.5 GB to about 2.1 GB, i.e., by a factor of 290.4. The feature-transformation time is also reduced from 21\,930 s to 231 s.

\begin{table}[H]
\centering
\caption{Comparison between the original CSF-based preprocessing and
the CSFG-based preprocessing with constant-orbital feature elimination. The CSF-based preprocessing followed the original Python
workflow~\cite{Bilous2023}, whereas the CSFG-based preprocessing used
32 MPI processes for parallel feature extraction with
\nolinkurl{nn_load_csfg_mpi}.}
\label{tab:feature_comparison}
\begin{tabular}{cccc}
\toprule
 & \makecell[c]{CSF-based} & \makecell[c]{CSFG-based with\\feature elimination} & \makecell[c]{Reduction\\ratio} \\
\midrule
Basis size & 437\,285\,546 & 15\,056\,337 & 29.0 \\
Number of orbitals & 115 & 46 & 2.5 \\
Total features (billion) & 150.86 & 2.08 & 72.6 \\
NumPy array (GB) & 603.5 & 2.1 & 290.4 \\
Execution time (s) & 21\,930 & 231 & 95.0 \\
\bottomrule
\end{tabular}
\end{table}

\subsection{\label{Partitioning of CSFGs - Importance labels}CSFG partitioning and importance labels}

During each iteration, the RCI calculation on the current CSFG set provides training labels for the NN. The CSFGs in the current set are partitioned and labeled as important or unimportant according to their weights. In the previous NN-supported pCI method~\cite{Bilous2024,Bilous2025}, calculations involving multiple target states are handled by assigning each relativistic configuration the largest weight among the target states. At each iteration, a user-specified cutoff \(x_i\) partitions the configurations into important and unimportant classes; decreasing \(x_i\) in later iterations allows configurations with smaller weights to be included.

In the present NN workflow, the normalized CSFG weight \(W_g\) defined in Eq.~\eqref{eq:csfg_weight} is used to generate the importance labels. This definition allows the selection to account for all target states in the same \(J^{\pi}\) block.

The weight distribution of CSFs, CSFGs, or relativistic configurations varies substantially across atomic systems and basis-set sizes. Therefore, manually choosing a cutoff sequence \(x_i\) is difficult. Here we adopt an adaptive cumulative truncation strategy inspired by the cumulative-weight idea described in Section~\ref{Condensation}, but use it to define NN training labels during the iterations rather than as a separate condensation step. After each RCI calculation, the CSFGs are sorted in descending order of \(W_g\) and accumulated until the cumulative sum reaches a preset fraction \(\delta_i\). The CSFGs contributing to this cumulative sum are labeled important, while the rest are labeled unimportant. The corresponding importance cutoff is determined as \(x_i = \log_{10} W_{g,\mathrm{cut}}\), where \(W_{g,\mathrm{cut}}\) is the weight of the last included CSFG. In the automated script, \(\delta_i\) is generated from \(1-10^{p_i}\), with the negative exponent \(p_i\) gradually decreased until the requested lower bound is reached. This makes the cutoff increasingly permissive as the selected set approaches convergence, while preventing uncontrolled inclusion of CSFGs with negligible weights. The calculation of \(W_g\), the cumulative summation, and the determination of \(x_i\) are performed by the {\sc Graspg} extension module \nolinkurl{rmixcsfg_gm}, which stores \(W_g\) in the binary file \nolinkurl{nameb.gwgt} and writes the current cutoff to \nolinkurl{cutoff.txt}.

Because additional random sampling is included (see Stage D in Section~\ref{Algorithm workflow}), CSFGs that have already been evaluated and found to have negligible weights should not be repeatedly included during the subsequent iterations. We therefore introduce a discard cutoff \(s_i\). Any CSFG with \(\log_{10} W_g < s_i\) is marked as discarded and is excluded from later NN predictions and RCI calculations. The value of \(s_i\) is set relative to \(x_i\), for example \(s_i=x_i-6\), so the discarded CSFGs have weights at least six orders of magnitude smaller than the current importance cutoff. This prevents wasteful recomputation and keeps the algorithm focused on potentially important CSFGs.

\subsection{\label{Algorithm workflow}Algorithm workflow}

\begin{figure}[t]
  \centering
  \includegraphics[width=0.98\textwidth]{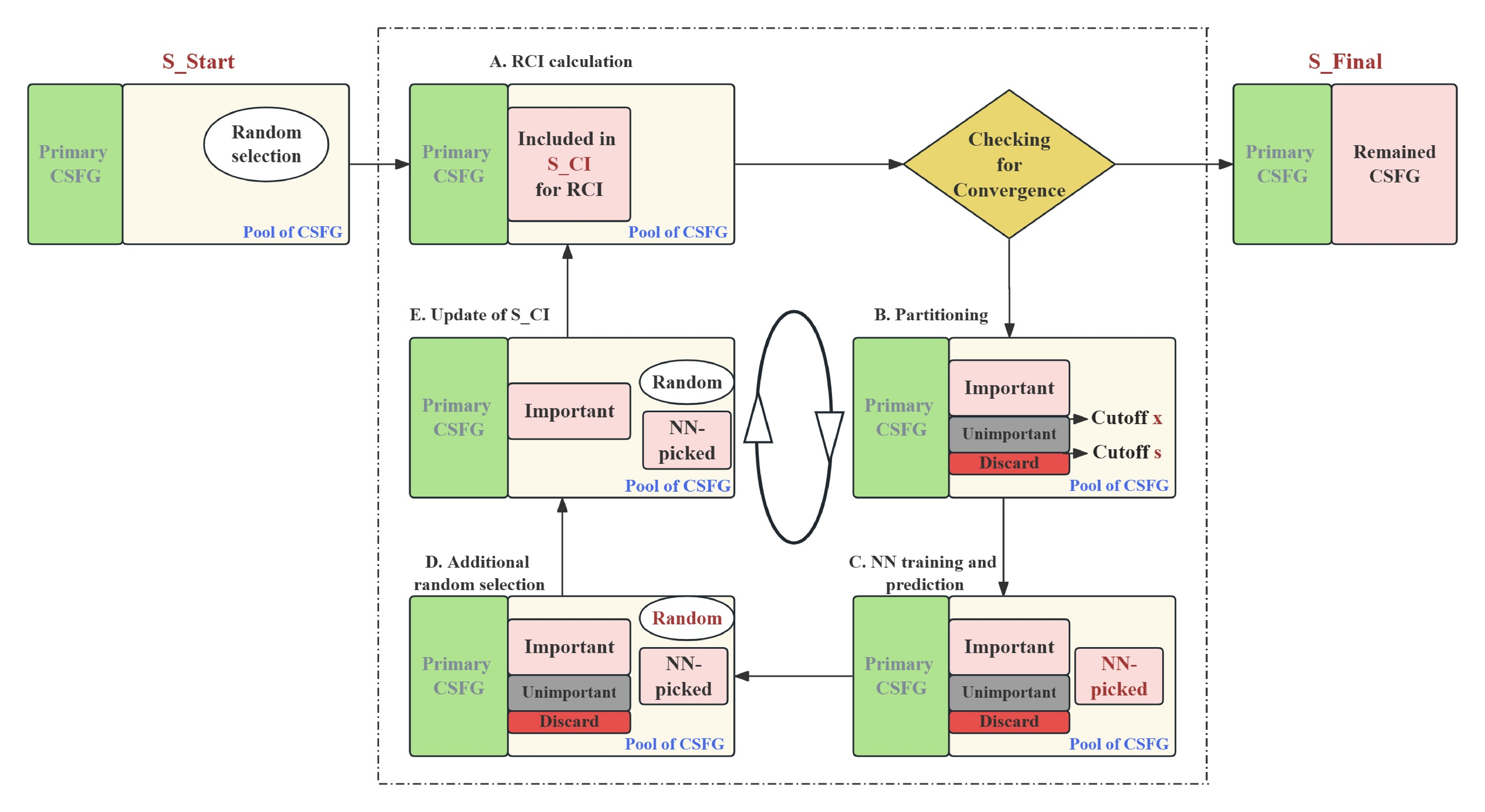}
  \caption{Flowchart of the CSFG-based NN selection workflow in {\sc Graspg}.}
  \label{fig:nn_graspg}
\end{figure}

The algorithm workflow is illustrated in Fig.~\ref{fig:nn_graspg}. Each NN basis-selection workflow is performed independently for a single \(J^{\pi}\) block. Within that block, the full CSFG set is divided into a primary CSFG set, which is retained throughout the NN iterations, and a sample pool subject to NN selection. The primary CSFG set is specified by the user according to the atomic system, target states, and adopted correlation model. Practical considerations for defining this set are discussed and exemplified in Section~\ref{Necessary calculations before the algorithm}. Given the predefined primary set, the {\sc Graspg} module \nolinkurl{rcsfgzerofirst_csfg} partitions and reorders the full CSFG set so that the primary CSFGs appear first, followed by the non-primary CSFGs. The workflow starts from an initial set \(\mathcal{S}_{\text{start}}\), consisting of all primary CSFGs and a fraction of the CSFGs randomly selected from the sample pool.

Each iteration consists of five stages, labeled A to E:

\begin{itemize}
    \item \textbf{Stage A (RCI calculation)}: Perform an RCI calculation on the current CSFG set \(\mathcal{S}_{\text{CI}}\) using the \nolinkurl{rci_block_csfg_mpi} program from the {\sc Graspg} package.
    
    \item \textbf{Stage B (Partitioning)}: Compute the normalized weight \(W_g\) for each CSFG  and the importance cutoff \(x_i\) using \nolinkurl{rmixcsfg_gm} according to cumulative fraction \(\delta_i\). 
    The non-primary CSFGs in \(\mathcal{S}_{\text{CI}}\) are then partitioned by the Python code into three categories: important, unimportant, and discarded. The partitioning is based on \(W_g\), \(x_i\), and the discard cutoff \(s_i\).
    
    \item \textbf{Stage C (NN training and prediction)}: Train the NN using the current RCI results to learn the features of important CSFGs. The trained NN is then applied to the remaining CSFGs not yet in \(\mathcal{S}_{\text{CI}}\) to predict potentially important CSFGs.
    
    \item \textbf{Stage D (Additional random selection)}: If the update relied only on NN predictions, important CSFGs that had never entered an RCI calculation could remain absent from the training set, causing the NN to miss their features and converge spuriously. To mitigate this risk, some CSFGs not selected by the NN are randomly selected and added to the next CSFG set \(\mathcal{S}_{\text{CI}}\). We define an adaptive random-exploration factor \(\alpha_i = 0.5 + 2.5 e^{-(i-1)/2}\) (with \(i\) starting from 1), such that the number of additionally sampled CSFGs is \(\alpha_i\) times the number of NN-predicted important CSFGs. The factor is large in early iterations to encourage exploration and decays exponentially as the training set becomes more informative.
    
    \item \textbf{Stage E (Update of \(\mathcal{S}_{\text{CI}}\))}: Remove the CSFGs below the current importance cutoff \(x_i\), keep the important CSFGs already identified, add the NN-predicted CSFGs and the CSFGs selected for random exploration, and mark the discarded CSFGs so that they are no longer considered in subsequent iterations. The resulting set becomes the next \(\mathcal{S}_{\text{CI}}\).
\end{itemize}

During the iterations, each RCI calculation provides the normalized weights \(W_g\) of the CSFGs in the current set \(\mathcal{S}_{\text{CI}}\), which are used to define the NN training labels and guide the update of the CSFG set. Because the goal of the iterative stage is to determine relative CSFG importance rather than final spectroscopic energies, higher-order relativistic effects such as Breit and QED corrections are omitted and only \(\hat{H}_{\mathrm{DC}}\) is included. The ZF method described in Section~\ref{Zero-First-Order Method} is used to accelerate these RCI calculations. The primary CSFGs form the \(P\) space and the remaining CSFGs in \(\mathcal{S}_{\text{CI}}\) form the \(Q\) space.

After each RCI calculation in Stage A, the energies from the current and previous iteration are compared. When the maximum difference in the absolute energies, \(\Delta E_{\text{max}}^{\text{abs}}\), across all target states falls below a specified tolerance (e.g., \(1\ \text{cm}^{-1}\)), the loop terminates.

Then a final cutoff is applied to \(\mathcal{S}_{\text{CI}}\) to remove the remaining unimportant CSFGs and generate the final CSFG set \(\mathcal{S}_{\text{final}}\), a compact subset of the full CSFG set. The compact CSF space generated by \(\mathcal{S}_{\text{final}}\) can then be used directly for the final all-relativistic RCI calculation, in which Breit and QED corrections are included.

%% forth section: Structure and installation of the package
\section{\label{Structure and installation}Structure and installation of the package}

\subsection{\label{structure}Structure of the package}

The package consists of a {\sc Graspg} extension patch and a Python-based NN toolkit:

{\fontfamily{cmr}\fontseries{bx}\fontshape{sc}\selectfont Graspg}
\textbf{extension patch}: The patch adds two Fortran-side modules required by the algorithm, \nolinkurl{nn_load_csfg_mpi} and \nolinkurl{rmixcsfg_gm}. The former prepares compact CSFG features for NN input, while the latter extracts CSFG weights from RCI results and determines the adaptive cumulative cutoff.

\textbf{Python NN toolkit}: The Python part is tightly coupled to the {\sc Graspg} workflow and contains three main modules:
\begin{itemize}
    \item \nolinkurl{Feature_Transform.py}, which converts the binary CSFG features into the pool-CSFG NumPy feature array;
    \item \nolinkurl{Iteration_manage.py}, which controls the initialization, iteration, and final-selection stages;
    \item \nolinkurl{nn_graspg_toolkit.py}, which implements memory-aware data loading, CNN training and prediction, CSFG marking, and {\sc Graspg} input-file generation.
\end{itemize}
A detailed description of the Python functions and usage is provided in Section~\ref{sec:running-graspg}. The toolkit uses NumPy~\cite{Harris2020} for array processing and TensorFlow~\cite{tensorflow2015-whitepaper} for the NN implementation. The versions of Python and the main libraries used in this work are listed in Table~\ref{tab:versions}. 

\begin{table}[htbp]
\centering
\caption{Python environment used in the test workflow.}
\begin{tabular}{ll}
\toprule
\textbf{Package} & \textbf{Version} \\
\midrule
Python & 3.12.4 \\
NumPy & 1.26.4 \\
TensorFlow & 2.16.1 \\
psutil & 5.9+ \\
\bottomrule
\end{tabular}
\label{tab:versions}
\end{table}

The package also provides the shell script \nolinkurl{NN_driven_GRASPG.sh} for executing the full workflow automatically, together with the example input files used in Section~\ref{sec:running-graspg}. Further operational details are given in \nolinkurl{README.pdf}, which is also included in the package.

\subsection{\label{sec:graspg-updates}Necessary updates to {\sc Graspg}}

The {\sc Graspg} patch adds the following two executable modules:

\begin{itemize}
    \item \nolinkurl{nn_load_csfg_mpi}: 
    This module reads the CSFG description file (\nolinkurl{name.g}) in parallel, removes constant-orbital feature channels, extracts the three feature streams describing orbital occupations and spin-angular coupling, and stores the compact CSFG features in binary format in \nolinkurl{compact_listb.dat}.
    \item \nolinkurl{rmixcsfg_gm}: This module reads the CSFG description files (\nolinkurl{name.g}, \nolinkurl{name.l}) and the RCI mixing-coefficient file (\nolinkurl{name.cm}), computes the normalized weight \(W_g\) for each CSFG, and stores the weights in binary format in \nolinkurl{nameb.gwgt}. It also determines the cumulative cutoff from the preset cumulative fraction \(\delta_i\) and writes \(x_i = \log_{10} W_{g,\mathrm{cut}}\) to \nolinkurl{cutoff.txt}.
\end{itemize}

These updates do not change the existing {\sc Graspg} workflow for conventional calculations; they only add the interfaces needed by the Python NN toolkit.

\subsection{\label{installation}Installation}
The CSFG-based NN basis-selection package can be installed as follows.

\textbf{1. Installation of {\fontfamily{cmr}\fontseries{bx}\fontshape{sc}\selectfont Graspg}:} {\sc Graspg}~\cite{Si2025} extends {\sc Grasp}2018~\cite{FroeseFischer2019} and requires an existing {\sc Grasp}2018 installation. Both packages are available on GitHub at \url{https://github.com/compas}. The {\sc Grasp}2018 installation should be completed first following its manual, and {\sc Graspg} can then be installed according to Ref.~\cite{Si2025}.

\textbf{2. Updating {\fontfamily{cmr}\fontseries{bx}\fontshape{sc}\selectfont Graspg}:} The extension modules described in Section~\ref{sec:graspg-updates} are provided as a patch file \nolinkurl{GRASPG-update.patch}. Copy this file to the \nolinkurl{srcg} directory of the {\sc Graspg} installation and apply the patch from that directory:

\begin{lstlisting}
>>patch -p1 < GRASPG-update.patch
\end{lstlisting}

Then recompile {\sc Graspg} using \texttt{make}. After successful compilation, the executables \nolinkurl{nn_load_csfg_mpi} and \nolinkurl{rmixcsfg_gm} will be generated in the \nolinkurl{bin} directory of the {\sc Grasp}2018/{\sc Graspg} installation.

\textbf{3. Configuring the Python environment:} The Python toolkit requires NumPy, TensorFlow, and psutil. The versions used in the test workflow are listed in Table~\ref{tab:versions}. Other Python 3.10--3.12 environments may also be used, provided that compatible versions of the required libraries are available on the target platform. The files in the Python toolkit can be copied to the working directory of a calculation.

The complete source code, {\sc Graspg} patch file, workflow scripts, README file, and example input files are provided as open-source supplementary material.

%% fifth section: NN-supported RCI with {\sc Graspg}
\section{\label{sec:running-graspg}Running the method with {\fontfamily{cmr}\fontseries{bx}\fontshape{sc}\selectfont {Graspg}}}

We demonstrate the method using the five lowest states of Ni\(^{12+}\) (\(Z=28\), \(N_{\text{e}}=16\), ground-state configuration [Ne]\(3s^2 3p^4\)). This ion is relevant to high-precision optical-clock studies and searches for new physics~\cite{Kozlov2018,Cheung2025}.

We first describe a moderate benchmark for which a direct full-space RCI calculation remains feasible and can therefore be used to monitor the convergence of the NN selection procedure. We then present both the manual step-by-step execution and the automated scripted workflow. In the discussion, we analyze the convergence, accuracy, and efficiency of the method. Finally, we apply the method to a larger-scale calculation for which direct full-space RCI is impractical, demonstrating how the method enables realistic large-scale high-precision calculations.

The detailed usage of the standard {\sc Grasp}2018 and {\sc Graspg} programs is not repeated here; the reader is referred to the original references and supplementary materials~\cite{Si2025,FroeseFischer2019}.

\subsection{\label{Necessary calculations before the algorithm}Preparatory calculations}

Before applying the CSFG-based NN selection method, we sequentially obtain the one-electron radial wave functions, define the primary CSFG set, and construct the full CSFG set for RCI.

The one-electron radial wave functions are obtained using the MCDHF method with the layer-by-layer optimization strategy~\cite{FroeseFischer2019}, employing the {\sc Graspg} modules \nolinkurl{rcsfggenerate_csfg}, \nolinkurl{rwfnestimate_csfg}, \nolinkurl{rangular_csfg_mpi}, and \nolinkurl{rmcdhf_csfg_mpi}. In the MCDHF calculations, the CSF space is generated by allowing SD excitations from the six valence electrons of the \(3s^2 3p^4\) reference configuration, thereby accounting for valence-valence (VV) correlation. To validate the NN selection procedure while keeping the corresponding direct full-space RCI calculation feasible, we restrict the active set to \(n\le 7\) and \(l\le 6\). The resulting radial wavefunction file is named \nolinkurl{Z28_ne16.w}.

With the radial orbitals fixed, the primary CSFG set is specified by the user according to the target states and adopted correlation model. It should contain the CSFGs needed to represent the dominant configurations and important correlation effects, as indicated by the MCDHF expansion, preliminary calculations, or known strong configuration mixing. The CSFGs corresponding to this MCDHF expansion are therefore adopted as the primary set and included in every NN iteration. The number of primary CSFGs is system dependent; the set should provide an adequate description of the target states while remaining computationally manageable.

Subsequently, a significantly larger CSF space is generated for RCI calculations under the same active set using the {\sc Graspg} module \nolinkurl{rcsfggenerate_csfg}. First, the single-reference configuration \(3s^2 3p^4\) is extended to the MR configuration set \(\{3s^2 3p^4,\allowbreak\;3s3p^4 3d,\allowbreak\;3s^2 3p^2 3d^2\}\). All 16 electrons in these MR configurations are then allowed to undergo SD excitations. This accounts for all possible valence-valence (VV), core-valence (CV), and core-core (CC) correlations and also includes potentially important triple and quadruple (TQ) excitations relative to \(3s^2 3p^4\). The CSFGs generated in this step constitute the full CSFG set for the RCI calculation.

The CSF spaces for different \(J^{\pi}\) blocks are generated separately. The numbers of primary CSFGs, full CSFGs, and generated CSFs are listed in Table~\ref{tab:csf_counts}. In the following, we illustrate the workflow using the two lowest states in the even \(J=0\) block (\(0^{+}\) block).

\begin{table}[htbp]
\centering
\caption{Numbers of primary CSFGs, full CSFGs, and generated CSFs for each \(J^{\pi}\) block in the Ni\(^{12+}\) benchmark calculation.}
\label{tab:csf_counts}
\begin{tabular}{cccc}
\toprule
\(J^{\pi}\) & Primary CSFGs & Full CSFGs & Generated CSFs \\
\midrule
\(0^{+}\) & 677 & 366\,746 & 2\,797\,018 \\
\(1^{+}\) & 1\,678 & 1\,049\,627 & 7\,959\,217 \\
\(2^{+}\) & 2\,553 & 1\,604\,630 & 11\,962\,361 \\
Total & 4\,908 & 3\,021\,003 & 22\,718\,596 \\
\bottomrule
\end{tabular}
\end{table}

As described in Section~\ref{Algorithm workflow}, the primary CSFGs form the \(P\) space and the remaining CSFGs in \(\mathcal{S}_{\text{CI}}\) form the \(Q\) space in the ZF method. The {\sc Graspg} module \nolinkurl{rcsfgzerofirst_csfg} is used to partition and reorder the full CSFG list so that the primary CSFGs forming the \(P\) space appear first, followed by the remaining CSFGs. The resulting full-space description files are renamed \nolinkurl{Z28_ne16.g}, \nolinkurl{Z28_ne16.l}, and \nolinkurl{Z28_ne16.icut}; the last file records the number of CSFGs in the \(P\) space. The primary CSFGs are also stored separately in \nolinkurl{Z28_ne16_primary.g}.

To provide an internal reference for the benchmark, we also perform an RCI calculation on the full CSFG set \nolinkurl{Z28_ne16.g}. This calculation uses the same settings as the intermediate iterations: the same \(P\) space is used in the ZF treatment, and only \(\hat{H}_{\mathrm{DC}}\) is included. The resulting energies are used only to monitor the convergence of the NN iterations and are labeled "reference RCI" in Table~\ref{tab:result_analysis}.

\subsection{\label{CSFG-based NN selection workflow}CSFG-based NN selection workflow}

All prepared {\sc Graspg} files and Python modules should be placed in the working directory. The required Python modules are \nolinkurl{Feature_Transform.py}, \nolinkurl{nn_graspg_toolkit.py}, and \nolinkurl{Iteration_manage.py}; their roles have been summarized in Section~\ref{structure}.

\nolinkurl{nn_load_csfg_mpi} and all RCI calculations in this section were performed using 32 MPI processes.

\subsubsection{\label{CSFG feature transforming}CSFG feature transformation}

First, \nolinkurl{nn_load_csfg_mpi} reads the CSFG description file, eliminates constant-orbital feature channels, and writes the compact integer-valued features to the binary file \nolinkurl{compact_listb.dat}: 

\begin{lstlisting}[breaklines]
>>mpirun -np 32 nn_load_csfg_mpi
 Welcome to program nn_load_csfg_mpi

 This MPI program reads <name>.g together with <name>.l,
    eliminates constant-orbital CSFG features, and outputs
    compact_listb.dat for the NN feature-transform step.
 If requested, compact_list.dat is also written in ASCII format.

 Name of state: 
>>Z28_ne16
 Output the files in ASCII format? (y/n)
>>n
 ...
 iblock =           1   NCSF(L) =        2546   NCSF(G) =      366746   NCSFs =     2797018
 ...
 Number of relativistic subshells: 49
 2*J + 1 =   1
 ...
 Collecting orbital-retention flags across MPI processes ...
 Generating compact CSFG feature arrays ...
 The following  33 orbitals are retained:
   1s    2s    2p-   2p    3s    3p-   3p    3d-   3d 
   6s    7s    6p-   7p-   6p    7p    6d-   7d-   6d 
   7d    6f-   7f-   6f    7f    6g-   7g-   6g    7g 
   6h-   7h-   6h    7h    7i-   7i 
 Total CSFGs:         366746, Number of blocks:    1
 Output compact_listb.dat in binary format ...
 ...
 Ratio of compact to full orbital-feature size:   0.6735
 ...
NN_LOAD_CSFG_MPI: Execution complete.
\end{lstlisting}

In this example, the CSFG file contains 49 relativistic subshells before feature elimination, while 33 subshell channels are used as the NN input. The feature channels associated with the \(4s\) -- \(5g\) correlation orbitals are constant over the CSFG pool and are therefore removed at this stage.

The binary file \nolinkurl{compact_listb.dat} is then processed by \nolinkurl{Feature_Transform.py}, which writes the pool-CSFG feature array \nolinkurl{Z28_ne16.npy} and records the numbers of primary and pool CSFGs in \nolinkurl{feature_transform.log}. Chunked processing and buffer reuse are used during this conversion to reduce memory use for large CSFG pools.

\begin{lstlisting}[breaklines]
>> python Feature_Transform.py
Number of electrons:
>> 16
Atomic number:
>> 28
Feature transformation started.
...
Primary and pool CSFG counts: Primary=677, Pool=366069
...
Pool CSFGs processed: 366069
...
Feature transformation completed.
...
\end{lstlisting}

Together, the {\sc Graspg}-side loading and the Python-side feature transformation reduce both preprocessing time and NumPy-array storage demand, as summarized in Table~\ref{tab:feature_comparison}.

\subsubsection{\label{NN setting}NN settings}

In the NN stage, each preprocessed CSFG is represented by an \(N_{\mathrm{orb}}^{\mathrm{ret}}\times 3\) integer-valued feature array, where the three channels correspond to occupation, subshell-angular, and intermediate-coupling descriptors. Here \(N_{\mathrm{orb}}^{\mathrm{ret}}\) denotes the number of orbitals retained after constant-orbital feature elimination. Following the CSF-feature CNN of Bilous \emph{et al.}~\cite{Bilous2023}, we use the same basic one-dimensional convolutional architecture, but apply it at the CSFG level. The model implemented in \nolinkurl{nn_graspg_toolkit.py} can be summarized as follows:

\begin{lstlisting}[breaklines]
Input: (N_orb, 3)
Conv1D(96, kernel_size=3, activation='relu')
Conv1D(16, kernel_size=1, activation='relu')
Flatten()
Dense(150, activation='relu')
Dense(120, activation='relu')
Dense(90, activation='relu')
Dense(2, activation='softmax')
Optimizer: Adam
Loss: categorical cross-entropy
\end{lstlisting}

The first convolutional layer scans neighboring positions in the ordered orbital-descriptor sequence and extracts local CSFG patterns. The second convolutional layer, with kernel size 1, combines the feature maps at each orbital position. The flattened representation is then passed through three dense layers, and the final softmax layer gives the probabilities of the two classes, important and unimportant.

The NN is trained with importance labels generated from the CSFG weights \(W_g\) and the adaptive cumulative cutoff described above. At iteration \(i\), CSFGs in the current training set with \(\log_{10}W_g \ge x_i\) are labeled as important, and the rest are labeled as unimportant. Using these two-class labels, the CNN is trained as a softmax classifier under the following training settings.

The maximum number of training epochs is set to 25, with a validation split of 0.2. For training sets with no more than 9600 samples, the batch size is set to 32; otherwise, it is chosen on a logarithmic scale according to the training-set size and capped at 131072. Three stopping criteria are used: the standard Keras early-stopping callback monitoring validation accuracy, a high-accuracy stopping condition, and a validation-loss progress monitor. These criteria reduce unnecessary training epochs while restoring the model parameters from the best validation epoch.

In the prediction stage, the trained NN is applied to the remaining pool CSFGs. To handle large CSFG pools, \nolinkurl{nn_graspg_toolkit.py} uses memory-aware loading of the NumPy feature array, memory mapping when necessary, pre-allocated buffers, and chunked prediction. For each candidate CSFG, the predicted probability of belonging to the important class is recorded. CSFGs with a predicted probability of at least 0.5 are treated as NN-predicted important CSFGs and are used in the next update step.

\subsubsection{\label{First RCI calculation}First RCI calculation}

The initial CSFG set \(\mathcal{S}_{\text{start}}\) and the untrained NN model are initialized using \nolinkurl{Iteration_manage.py}, together with the files required for the subsequent NN iterations:

\begin{lstlisting}[breaklines]
>> python Iteration_manage.py
TensorFlow thread settings have been applied.
Number of electrons:
>> 16
Atomic number:
>> 28
Select Algorithm Stage (0=Start, 1=Iteration, 2=Final):
>> 0
Fraction of random selection from pool CSFGs (e.g., 0.01-0.05):
>> 0.05
...
Step 1: Check NumPy array memory requirements.
...
Step 2: Initialize NN-GRASPG toolkit.
...
NN-GRASPG toolkit initialization completed.
Full sample pool contains 366069 CSFGs
Feature parameters per CSFG: 99
Number of primary CSFGs: 677
Number of pool CSFGs: 366069
...
Step 3: Initialize algorithm state.
...
Initializing marking and weight arrays...
...
Initializing the NN setup...
...
Step 4: Generate the initial CSFG set S_Start for the first RCI calculation.
...
Random-selection fraction: 0.05
Initial pool CSFGs selected: 18303
...
The initial set S_Start contains 677 primary CSFGs and 18303 pool CSFGs.
...
Saving algorithm state (arrays and NN model).
...
Iteration_manage.py completed.
\end{lstlisting}

After this initialization, the description files of \(\mathcal{S}_{\text{start}}\), namely \nolinkurl{Z28_ne16_0.g} and \nolinkurl{Z28_ne16_0.l}, are written together with the radial wavefunction file \nolinkurl{Z28_ne16_0.w}. These files provide the input for the first RCI calculation. The log file \nolinkurl{Z28_ne16_0.log} records the main parameters and program status, while intermediate data such as CSFG marking arrays and NN-related configuration files are stored in \nolinkurl{NN_Saved_Iter_0}.

The first RCI calculation is then performed with \nolinkurl{rci_block_csfg_mpi}, using \(\hat{H}_{\mathrm{DC}}\) together with the ZF method. This calculation provides the first set of mixing coefficients from which the CSFG weights are evaluated. The resulting energy levels are extracted using \nolinkurl{rlevels} and saved in \nolinkurl{Z28_ne16_0.clev}. These data serve as the starting point of the iterative selection loop.

\subsubsection{\label{Iteration stage}Iteration stage}

For the \(i\)-th iteration (\(i \ge 1\)), \nolinkurl{rmixcsfg_gm} reads the RCI mixing coefficients from iteration \(i-1\), evaluates the normalized weights \(W_g\) of CSFGs, and determines the cumulative weight cutoff using \(\delta_i\):
\begin{equation}
\begin{aligned}
p_i &= \max\left(p_{\mathrm{init}} + (i-1)p_{\mathrm{step}},\, p_{\mathrm{max}}\right),\\
\delta_i &= 1 - 10^{p_i}.
\end{aligned}
\label{eq:delta_i}
\end{equation}

In the benchmark discussed below, \(p_{\mathrm{init}}=-6\), \(p_{\mathrm{step}}=-0.5\), and \(p_{\mathrm{max}}=-9\), so that the target cumulative weight fraction increases from \(\delta_1=0.999999\) toward \(1-10^{-9}\) over the subsequent iterations, while the corresponding CSFG-weight cutoff gradually shifts toward lower weights. The execution of \nolinkurl{rmixcsfg_gm} is as follows:

\begin{lstlisting}[breaklines]
>> rmixcsfg_gm
 Welcome to program rmixcsfg_gm
 
 This program reads <name>.m or <name>.cm, outputs <name>.gm, <name>b.gwgt and cutoff.txt
 The NN program will read the CSFG weights from <name>b.gwgt.

 Name of state: 
>> Z28_ne16_0 # Z28_ne16_{i-1}
 Expansion coefficients resulting from CI calculation (y/n)?
>> y
 Input c2evecblklim, to obtain the possible cutoff factor:
>> 0.999999 # delta_i
 ...
 Number of CSFGs =       18980
 -8.90E+00    1.26E-09 = Log10(epsmin), epsmin, CSFG-NN
 ...
STOP Normal Exit
\end{lstlisting}

The printed value \(\log_{10}(\text{epsmin})\) is used as the importance cutoff \(x_i=\log_{10}W_{\mathrm{cut}}^{(i)}\), and is also stored in \nolinkurl{cutoff.txt}. The CSFG weights  \(W_g\) for the current RCI set are written in binary format to \nolinkurl{Z28_ne16_{i-1}b.gwgt}.

Next, \nolinkurl{Iteration_manage.py} reads \nolinkurl{Z28_ne16_{i-1}b.gwgt}, labels the current training CSFGs according to \(\log_{10}W_g \ge x_i\), trains the NN, and applies the trained model to the remaining pool CSFGs. The next \(\mathcal{S}_{\text{CI}}\) is then constructed by retaining the CSFGs above the current cutoff and adding both the NN-predicted important CSFGs and a small set of additionally sampled CSFGs:

\begin{lstlisting}[breaklines]
>> python Iteration_manage.py
TensorFlow thread settings have been applied.
Number of electrons:
>> 16
Atomic number:
>> 28
Select Algorithm Stage (0=Start, 1=Iteration, 2=Final):
>> 1
Current iteration number:
>> 1 # i-th iteration
Current cutoff (x_i):
>> -8.90 # x_i
Diff between discarded cutoff s_i and x_i (s_i - x_i):
>> -6
Random-exploration factor:
>> 3.0 # alpha_i
...
Step 1: Check NumPy array memory requirements.
...
Step 2: Initialize NN-GRASPG toolkit.
...
Step 3: Load algorithm state (arrays and NN model).
...
Step 4: Generate the next CSFG set S_CI using NN training and prediction.
...
Step 4.1: Read CSFG weights from the previous RCI calculation.
Total CSFGs in the previous RCI calculation (primary + pool): 18980
Number of pool-CSFG weights: 18303
...
Step 4.2: Train the NN from CSFG weights and cutoff.
...
CSFGs used for NN training: 18303
  of which important: 1762
...
Starting NN model training...
...
AutoStopTrainingProgressCallback triggered: improvement=-8.80e-03
NN was trained on 18303 samples
  training time: 23.47 seconds
  training device: CPU.
...
Step 4.3: Apply the NN to the pool CSFGs.
...
NN prediction completed.
Processed candidate pool CSFGs: 347766
Predicted important CSFGs: 11283
...
Step 4.4: Add randomly selected CSFGs and prepare GRASPG input files.
...
Random-exploration factor: 3.00, predicted important CSFGs: 11283
Additional randomly selected CSFGs: 33849
...
Step 4 completed.
...
Marked for training in next iteration: 45132
Marked for application in next iteration: 315597
Total discarded CSFGs: 5340
...
Ready for the next RCI calculation.
The new S_CI set contains 677 primary CSFGs and 45132 pool CSFGs.
...
Iteration_manage.py completed.
\end{lstlisting}

Here \(s_i=x_i-6\) is used as the discard cutoff for very small-weight CSFGs, while \(\alpha_i = 0.5 + 2.5 e^{-(i-1)/2}\) is the adaptive random-exploration factor. Similar to the initialization stage, this run generates a new set of files \nolinkurl{Z28_ne16_{i}.*} and a new data directory \nolinkurl{NN_Saved_Iter_{i}}. A subsequent RCI calculation is then performed on the updated \(\mathcal{S}_{\text{CI}}\) to obtain the current energy levels and mixing coefficients.

After each RCI calculation, convergence is checked from the change in the energies of the target states between two successive iterations. If the maximum difference \(\Delta E_{\text{max}}^{\text{abs}} < \epsilon\) (e.g., \(\epsilon = 1\ \text{cm}^{-1}\)), the iteration loop terminates and the procedure proceeds to the generation of the final CSFG set \(\mathcal{S}_{\text{final}}\). Otherwise, the iteration loop continues.

\subsubsection{\label{Final set generation}Final set generation}

After the iterative convergence criterion is satisfied, a final CSFG-weight cutoff is applied to the last \(\mathcal{S}_{\text{CI}}\) to form the final CSFG set \(\mathcal{S}_{\text{final}}\). In the benchmark, this step is performed after the eighth iteration. The final cumulative fraction is chosen according to the target accuracy and it is set to \(\delta_{\text{final}}=0.9999999\) here. Then \nolinkurl{rmixcsfg_gm} is used with the last RCI result to determine the corresponding cutoff \(x_{\text{final}}\).

Then \nolinkurl{Iteration_manage.py} constructs \(\mathcal{S}_{\text{final}}\):

\begin{lstlisting}
>> python Iteration_manage.py
TensorFlow thread settings have been applied.
Number of electrons:
>> 16
Atomic number:
>> 28
Select Algorithm Stage (0=Start, 1=Iteration, 2=Final):
>> 2
Final iteration number:
>> 8
Final cutoff (x_Final):
>> -11.1
...
Step 1: Check NumPy array memory requirements.
...
Step 2: Initialize NN-GRASPG toolkit.
...
Step 3: Load algorithm state (arrays and NN model).
...
Step 4: Generate the final CSFG set S_Final from the last RCI result and final cutoff.
...
Step 4.1: Read CSFG weights from the previous RCI calculation.
Total CSFGs in the previous RCI calculation (primary + pool): 114001
Number of pool-CSFG weights: 113324
...
Step 4.2: Apply final cutoff and prepare GRASPG input files.
...
Final CSFG set generation completed.
...
Ready for the final RCI calculation.
The final CSFG set S_Final contains 677 primary CSFGs and 57905 pool CSFGs.
...
Iteration_manage.py completed.
\end{lstlisting}

This run generates the CSFG description files \nolinkurl{Z28_ne16_Fi.g} and \nolinkurl{Z28_ne16_Fi.l} for \(\mathcal{S}_{\text{final}}\).

The compact CSF space generated by \(\mathcal{S}_{\text{final}}\) is then used for the final RCI calculation, in which Breit and QED corrections are included. In the benchmark analysis, these results are labeled "final set" in Table~\ref{tab:result_analysis} and are compared with the all-relativistic RCI results obtained from the full CSFG set, labeled "full set".

\subsubsection{\label{script}Run via script}

The script \nolinkurl{NN_driven_GRASPG.sh} executes the same workflow automatically, starting from the CSFG feature transformation described in Section~\ref{CSFG feature transforming}.

\begin{lstlisting}
>> ./NN_driven_GRASPG.sh 16 28 0.05 -6 -0.5 -9 -6 32 1.0 -7
\end{lstlisting}

The input parameters are, in order: number of electrons, atomic number, initial random-selection fraction, \(p_{\mathrm{init}}\), \(p_{\mathrm{step}}\), \(p_{\mathrm{max}}\) in Eq.~\ref{eq:delta_i}, the difference \(s_i-x_i\), number of parallel processes used in the RCI calculations, convergence tolerance in cm\(^{-1}\), and the exponent \(p_{\text{final}}\) defining \(\delta_{\text{final}}=1-10^{p_{\text{final}}}\).

The same procedure is applied to the \(1^{+}\) and \(2^{+}\) blocks using the same set of parameters. It is found that all the calculations converge after eight iterations. 

\subsection{\label{discussion of results}Discussion of results}

\subsubsection{\label{convergence and efficiency}Convergence and efficiency}

We first examine the convergence and efficiency of the CSFG-based NN selection method. Table~\ref{tab:result_analysis} reports the differences between the absolute energies of the target states obtained from the iterative \(\mathcal{S}_{\text{CI}}\) calculations and those from the reference RCI calculations described in Section~\ref{Necessary calculations before the algorithm}. The numbers of CSFGs and CSFs, together with the RCI wall times for the representative \(0^{+}\) block, are also listed to quantify the reduction in computational demand.

The initial set \(\mathcal{S}_{\text{start}}\) consists of the primary CSFGs and a fraction of CSFGs randomly selected from the sample pool. The energies obtained with this initial set differ from the corresponding reference RCI energies by approximately \(7\times10^{4}~\text{cm}^{-1}\), indicating that many important CSFGs are still absent. As the iterations proceed, \(\mathcal{S}_{\text{CI}}\) is progressively refined using NN predictions learned from the importance labels, and the energy differences decrease rapidly. After eight iterations, the energy-convergence criterion is satisfied, and the energies of all five target states agree with the corresponding reference RCI energies to within the target precision \(\varepsilon=1~\text{cm}^{-1}\). For the \(0^{+}\) block, \(\mathcal{S}_{\text{final}}\) contains 58\,582 CSFGs and 547\,352 CSFs, compared with 366\,746 CSFGs and 2\,797\,018 CSFs in the full set. The numbers of CSFGs and CSFs are therefore reduced by factors of about 6.3 and 5.1, respectively. Despite this substantial reduction, the final-set excitation energies agree with the full-set values within a few \(\mathrm{cm}^{-1}\), indicating that \(\mathcal{S}_{\text{final}}\) preserves the correlation contributions relevant at this accuracy scale.

\begin{sidewaystable}
\centering
\scriptsize
\setlength{\tabcolsep}{3pt}
\caption{Energy convergence and resource demand of the CSFG-based NN-selection method. During the iterations, only the Dirac-Coulomb Hamiltonian \(H_{\mathrm{DC}}\) is included and the ZF method is adopted. "All-relativistic RCI" refers to the RCI calculations on the final set and the full set, where Breit and QED corrections are included. The resource rows present the numbers of CSFGs/CSFs and the RCI wall times for the \(0^{+}\) block calculations.}
\label{tab:result_analysis}
\begin{tabular}{c *{12}{c}}
\toprule
\multirow{2}{*}{Level} &\multirow{2}{*}{\shortstack{Reference RCI\\ (Hartree)}} & \multicolumn{9}{c} {\(E_{\mathrm{abs}}^{\text{iter}} - E_{\mathrm{abs}}^{\text{ref}}\) (cm$^{-1}$)} & \multicolumn{2}{c}{All-relativistic RCI (cm$^{-1}$)}\\
\cmidrule(lr){3-11}\cmidrule(lr){12-13}
 & & 0 & 1 & 2 & 3 & 4 & 5 & 6 & 7 & 8 & Final set & Full set \\
\midrule
\(3s^2 3p^4\, {}^3P_2\) & -1458.7647995 & 76563.99 & 6315.93 & 287.84 & 60.20 & 18.46 & 5.60 & 1.87 & 0.50 & \textbf{0.26} & 0 & 0 \\
\(3s^2 3p^4\, {}^3P_1\) & -1458.6742447 & 77148.87 & 6876.23 & 923.59 & 195.31 & 34.72 & 5.73 & 1.69 & 0.42 & \textbf{0.24} & 19520.92 & 19518.51 \\
\(3s^2 3p^4\, {}^3P_0\) & -1458.6710945 & 74771.65 & 4751.23 & 664.35 & 134.36 & 42.16 & 11.76 & 2.96 & 0.81 & \textbf{0.37} & 20203.38 & 20197.59 \\
\(3s^2 3p^4\, {}^1D_2\) & -1458.5483318 & 78640.99 & 8712.53 & 309.39 & 70.17 & 21.07 & 6.54 & 2.22 & 0.57 & \textbf{0.31} & 47179.94 & 47179.12 \\
\(3s^2 3p^4\, {}^1S_0\) & -1458.3099099 & 75333.90 & 3201.67 & 443.14 & 86.89 & 28.18 & 8.60 & 2.41 & 0.66 & \textbf{0.29} & 99356.99 & 99352.35 \\
\midrule
%\multicolumn{13}{c}{Other items of \(0^{+}\) block}\\
\multicolumn{13}{c}{Number of CSFs/CSFGs and the RCI wall times of \(0^{+}\) block}\\
\midrule
\multicolumn{2}{c}{Number of CSFGs} & 18\,980 & 45\,809 & 33\,349 & 56\,984 & 65\,803 & 84\,311 & 98\,945 & 117\,121 & 114\,001 & 58\,582 & 366\,746 \\
\multicolumn{2}{c}{Number of CSFs} & 139\,095 & 350\,778 & 268\,301 & 479\,207 & 579\,402 & 757\,449 & 890\,771 & 1\,070\,279 & 1\,053\,561 & 547\,352 & 2\,797\,018 \\
\multicolumn{2}{c}{\multirow{2}{*}{RCI wall time (s)}} & 3 & 6 & 5 & 9 & 11 & 13 & 15 & 18 & 15 & 110 & \multirow{2}{*}{\textbf{1\,546}}\\
\cmidrule(lr){3-12}
 & & \multicolumn{9}{c}{Total: \textbf{377} (with 172 seconds additional time including NN)}\\
\bottomrule
\end{tabular}
\end{sidewaystable}

\begin{figure}[t]
  \centering
  \includegraphics[width=0.95\textwidth]{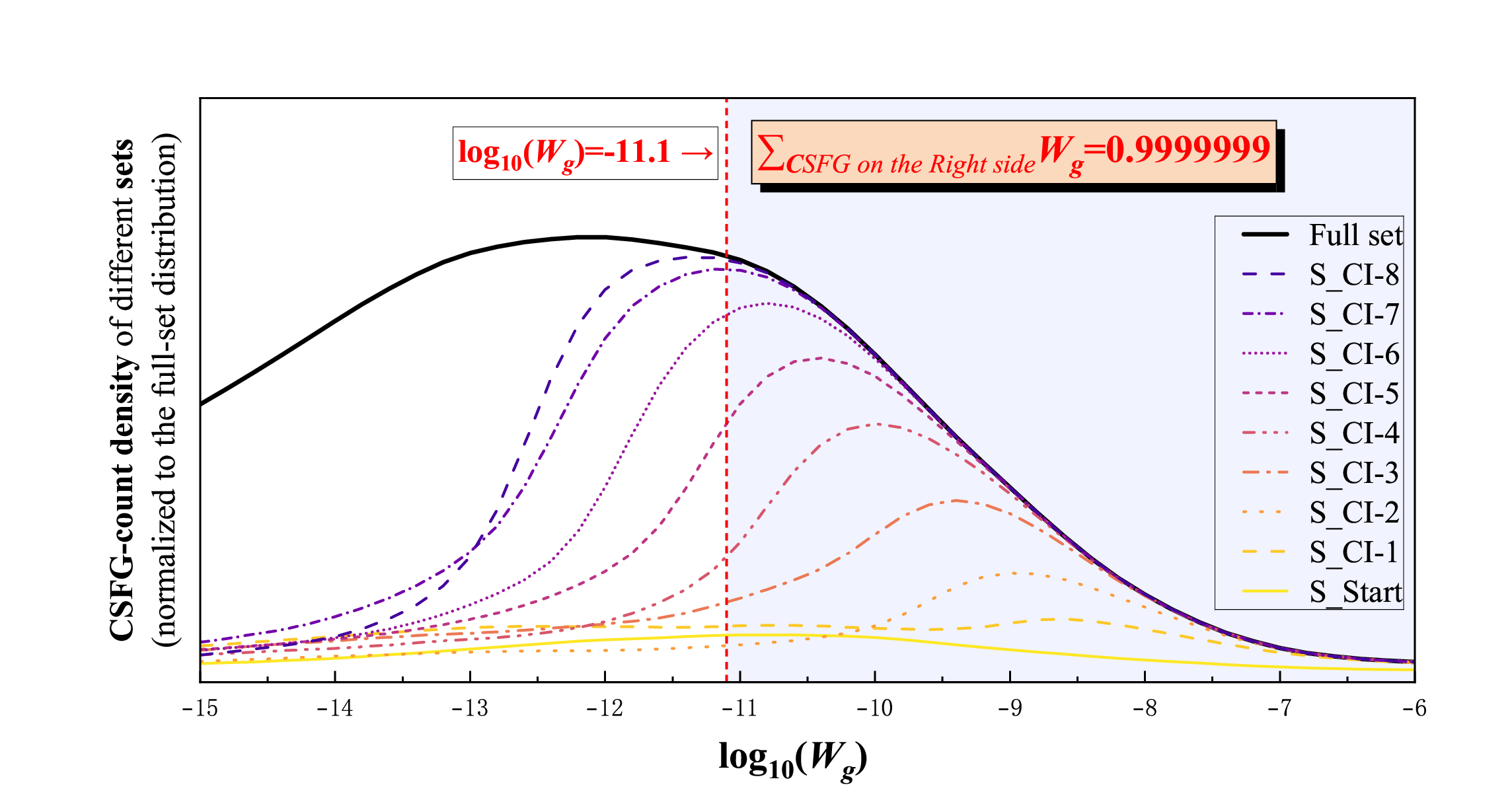}
  \caption{CSFG-count distributions of the full set and \(\mathcal{S}_{\mathrm{CI}}\) versus \(\log_{10}W_g\) for the \(0^{+}\) block of Ni\(^{12+}\). All curves are plotted on the same density scale normalized to the full-set distribution.}
  \label{fig:distribution}
\end{figure}

To visualize the iterative process more directly, Fig.~\ref{fig:distribution} shows the CSFG-count distributions as functions of \(\log_{10}W_g\) for the representative \(0^{+}\) block. The thick black curve represents the full CSFG set, while the colored curves show the CSFGs included in \(\mathcal{S}_{\text{CI}}\) during successive iterations, binned and smoothed according to their weights in the full set. All curves are plotted on the same density scale normalized to the full-set distribution. The red dashed line marks the cumulative cutoff determined from the full-set weight distribution for \(\delta_{\text{final}}=0.9999999\); CSFGs to the right of this cutoff carry this fraction of the total full-set weight. As the iterations proceed, the CSFGs included in \(\mathcal{S}_{\text{CI}}\) extend from the dominant high-weight region to progressively lower-weight regions. Because each update of \(\mathcal{S}_{\text{CI}}\) is mainly guided by NN predictions, this behavior shows that the NN selection expands \(\mathcal{S}_{\text{CI}}\) preferentially toward CSFGs that are important for the target ASFs, rather than simply increasing the set size uniformly. By the eighth iteration, nearly all CSFGs in the important region to the right of the red dashed line have been included.

Table~\ref{tab:result_analysis} also reports the RCI wall times for the \(0^{+}\) block. The complete CSFG-NN workflow takes 377 seconds in total, including the iterative RCI runs, the final-set RCI run, and 172 seconds spent on NN training, prediction, and workflow management. In contrast, the full-set RCI calculation requires 1546 seconds, corresponding to a wall-time reduction of 75.6\%. 
The peak memory required by a single RCI calculation is also reduced by a factor of 10.1, from 168.23 GB for the full-set RCI calculation to 16.66 GB for the final-set RCI calculation.
Since weakly contributing CSFGs occupy an increasing fraction of the basis set as the CSF space expands, these computational savings are expected to become more significant for larger-scale calculations.

Overall, the Ni\(^{12+}\) benchmark demonstrates that the CSFG-based NN selection method can reproduce the full-set results with a much smaller CSFG set. The comparison with the full-set calculation shows that the iterative selection retains the CSFGs carrying the dominant contributions to the target ASFs while reducing both the total wall time and the peak memory usage of the RCI calculations. These benchmark calculations support the effectiveness and reliability of the present NN selection method, which is therefore applied to a larger-scale calculation in the following section.

\subsubsection{\label{Application to realistic problem}Application to a larger-scale calculation}

In the preceding calculations, to keep the full-set RCI calculations computationally feasible, we restricted the active orbital set to \(n\le 7\) and \(l\le 6\). This restricted active set is suitable for validating the NN selection procedure, but it is not sufficient for high-precision calculations. For the larger-scale calculation, the active orbital set is therefore enlarged layer by layer up to \(n\le 11\) and \(l\le 8\).

The layer-by-layer MCDHF optimization showed that the target energies had converged by the \(n=11\) layer within the adopted VV orbital-optimization model. This provides a reliable one-electron radial basis for the subsequent RCI calculation. This follows the standard MCDHF--RCI procedure, in which active-set convergence is monitored at the MCDHF stage and additional electron-correlation effects are subsequently recovered by enlarging the CSF expansion in the RCI calculation~\cite{Li2025}. Therefore, for the present calculation, the computational challenge addressed is the size of this enlarged RCI expansion. 

For the subsequent RCI calculation, the MR set is formed by allowing the 6 valence electrons of \(3s^2 3p^4\) to be SD excited to the \(4s\), \(4p\), and \(3d\) orbitals. The CSFGs corresponding to these MR configurations are used as the primary CSFG set. The full CSFG set is then constructed by allowing SD excitations of all 16 electrons in the MR configurations to the enlarged active set. For the three \(J^{\pi}\) blocks considered below, the corresponding full CSF spaces for the RCI calculations contain \(1.51\times10^{8}\), \(4.37\times10^{8}\) and \(6.81\times10^{8}\) CSFs. 

The CSFG-based NN selection method was then applied independently to the \(0^{+}\), \(1^{+}\), and \(2^{+}\) blocks of the larger-scale calculation. The automated workflow was launched as
\begin{lstlisting}
>> ./NN_driven_GRASPG.sh 16 28 0.03 -6 -0.5 -9 -6 64 1.0 -6
\end{lstlisting}
Compared with the benchmark calculation, where the final cumulative fraction was set to \(\delta_{\mathrm{final}}=1-10^{-7}\) for a stringent comparison with the full-set RCI result, the larger-scale calculation used \(\delta_{\mathrm{final}}=1-10^{-6}\). This cutoff was chosen as a practical compromise between capturing the dominant CSFG contributions and keeping the final all-relativistic RCI calculations within the available computational resources.

Table~\ref{tab:csf_ultra_scale} summarizes the reduction achieved by CSFG-based NN selection for the present large-scale calculation. The final CSFG sets generate \(1.72\times10^{6}\), \(6.09\times10^{6}\), and \(1.31\times10^{7}\) CSFs for the \(0^{+}\), \(1^{+}\), and \(2^{+}\) blocks, respectively. These correspond to retained fractions of 1.1\%, 1.4\%, and 1.9\%.

\begin{table}[htbp]
\centering
\caption{Comparison of the number of CSFs before and after CSFG-based NN selection for the present large-scale calculations.}
\label{tab:csf_ultra_scale}
\setlength{\tabcolsep}{4pt}
\begin{tabular}{c c c c}
\toprule
\(J^{\pi}\) & \makecell[c]{Full CSF\\space} & \makecell[c]{Compact CSF space\\after NN selection} & \makecell[c]{Retained\\fraction} \\
\midrule
\(0^{+}\) & 150\,890\,394 & 1\,721\,696 & 1.1\% \\
\(1^{+}\) & 437\,285\,546 & 6\,091\,214 & 1.4\% \\
\(2^{+}\) & 680\,555\,087 & 13\,089\,907 & 1.9\% \\
\bottomrule
\end{tabular}
\end{table}

The final RCI calculations including Breit and QED corrections were then performed in the compact CSF spaces generated by the final CSFG sets. For the \(0^{+}\) and \(1^{+}\) blocks, the calculations were run with 64 MPI processes. For the \(2^{+}\) block, which has the largest compact CSF space, the {\sc Graspg} restart workflow was used to make the diagonalization feasible within the available memory. The Hamiltonian matrix elements were first constructed by \nolinkurl{rci_block_csfg_mpi} with 64 MPI processes and written to the \nolinkurl{rci.res} restart files. These files were then redistributed by \nolinkurl{rdistHmatrix_csfg} from the 64-process matrix-construction layout to a 4-process layout for diagonalization, after which \nolinkurl{rci_block_csfg_mpi} was rerun in restart mode to diagonalize the Hamiltonian.

\begin{table}[htbp]
\centering
\caption{Excitation energies (in cm\(^{-1}\)) relative to the
\(3s^2 3p^4\,{}^3P_2\) ground level from experiment and different
theoretical methods. The columns labeled ``diff'' give theory minus
experiment in cm\(^{-1}\).}
\label{tab:Ni_energies}
\small
\renewcommand{\arraystretch}{1.12}
\setlength{\tabcolsep}{2pt}
\begin{tabular*}{\linewidth}{
@{\extracolsep{\fill}}
c c c c c c c c c c
@{}
}
\toprule
\multirow{2}{*}{Level}
& \multirow{2}{*}{Exp.$^{a}$}
& \multicolumn{2}{c}{This work}
& \multicolumn{2}{c}{NN-pCI$^{b}$}
& \multicolumn{2}{c}{CI+all-order$^{c}$}
& \multicolumn{2}{c}{16-electron CI$^{c}$} \\
\cmidrule(lr){3-4} \cmidrule(lr){5-6}
\cmidrule(lr){7-8} \cmidrule(lr){9-10}
& & \(E\) & diff & \(E\) & diff & \(E\) & diff & \(E\) & diff \\
\midrule
\(3s^2 3p^4\,{}^3P_1\)
& 19542 & 19553 & 11  & 19408 & \(-134\) & 19547 & 5  & 19550 & 8 \\
\(3s^2 3p^4\,{}^3P_0\)
& 20079 & 20108 & 29  & 20038 & \(-41\)  & 20086 & 7  & 20081 & 2 \\
\(3s^2 3p^4\,{}^1D_2\)
& 47033 & 47124 & 91  & 47176 & 143      & 47063 & 30 & 47051 & 18 \\
\(3s^2 3p^4\,{}^1S_0\)
& 97836 & 97911 & 75  & 98149 & 313      & 97894 & 58 & 97771 & \(-65\) \\
\bottomrule
\end{tabular*}
\vspace{1.5mm}
\begin{minipage}{0.96\linewidth}
\centering
\footnotesize
\(^{a}\) Reference~\cite{NIST_ASD}, except for
\(3s^2 3p^4\,{}^3P_0\), for which the measured value of
Ref.~\cite{Cheung2025} is used;\\
\(^{b}\) Reference~\cite{Bilous2024};
\qquad
\(^{c}\) Reference~\cite{Cheung2025}.
\end{minipage}
\end{table}

Table~\ref{tab:Ni_energies} compares the final excitation energies obtained by the present method with experimental values from the NIST database~\cite{NIST_ASD} and with recent theoretical results from NN-pCI, CI+all-order, and 16-electron CI calculations~\cite{Bilous2024,Cheung2025}. The deviations of the present results from experiment are below \(0.2\%\), placing them on the same accuracy scale as the other theoretical calculations listed in the table. This agreement suggests that the compact CSF spaces generated by \(\mathcal{S}_{\text{final}}\) capture the dominant correlation effects relevant to the target ASFs considered here.

The comparison should also be viewed in terms of active-space size and computational cost. In the present calculation, the active orbital set extends to \(n\le 11\) and \(l\le 8\) (\(11l\)), producing a total full CSF space of \(1.27\times10^{9}\) CSFs over the three \(J^{\pi}\) blocks. The common configuration space underlying these blocks corresponds to approximately \(1.18\times10^{11}\) Slater determinants. After CSFG-based NN selection, only 1.1\%--1.9\% of the full-space CSFs are retained, and the complete large-scale workflow can be finished on a single CPU server in 10.1 h within a 2 TB memory capacity. For comparison, the Ni\(^{12+}\) NN-pCI calculation of Bilous \emph{et al.} used an active space up to \(22spdfgh20ikl\), comprising \(8.63\times10^{5}\) relativistic configurations and \(2.09\times10^{8}\) Slater determinants. The largest listed aggregate memory allocation was about 20 TB, and the CI runs required 71.2 h in total~\cite{Bilous2024}. In the CI+all-order calculation of Cheung \emph{et al.}, the initial \(17spdfg\) expansion was extended to \(n=29\) for the \(spdfg\) waves and to \(n=28\) for the \(h\) and \(i\) waves, with more than \(5\times10^{6}\) configurations treated across separate calculations. Their 16-electron CI calculation included outer-electron excitations up to \(20spdfghikl\) and inner-shell excitations through the \(spdfg\) waves; contributions from \(9.6\times10^{6}\) configurations were evaluated in 54 separate computations under a 31 TB memory limit~\cite{Cheung2025}. Thus, although the pCI-based values are closer to experiment for some levels, they were obtained through substantially more resource-intensive computational workflows. The present CSFG-based NN selection provides a lower-cost route to large-scale RCI calculations within the {\sc Graspg} framework while achieving a comparable level of accuracy.

The present large-scale calculation should be viewed as a practical operating point rather than the ultimate accuracy limit of the present CSFG-based NN selection method. Higher accuracy could be pursued by further extending the MR and active orbital spaces, at the cost of increased RCI dimensions and computational resources. This controllable tradeoff between accuracy and computational cost is the main purpose of the CSFG-based NN selection strategy.

\section{\label{Summary}Summary and outlook}
We have presented a CSFG-based NN basis-selection method and integrated it with the {\sc Graspg} package. The method shifts the NN selection unit from individual CSFs to CSFGs, removes constant-orbital descriptors before NN training, and determines importance cutoffs through adaptive cumulative truncation rather than fixed empirical thresholds. Initial random sampling and additional random sampling are included to keep the iterative training set exploratory and to reduce the risk of premature convergence to an incomplete CSFG set. The source code and example inputs are provided as open-source supplementary material.

Using Ni\(^{12+}\) as an example, the final CSFG set in the benchmark calculation reproduces the direct full-space all-relativistic RCI results at the few-\(\mathrm{cm}^{-1}\) level. For the representative \(0^{+}\) block, the complete workflow reduces the wall time by 75.6\%, and the peak memory required by a single RCI calculation is reduced by a factor of 10.1. In the larger-scale calculation, where direct full-space RCI is impractical, the method retains only 1.1\%--1.9\% of the full-space CSFs while yielding energies in agreement with experimental data and other resource-intensive theoretical calculations, with a substantially lighter computational workflow than the pCI calculations used for comparison. 

This work provides an effective solution for balancing the accurate description of electron correlation against computational cost, making it possible to perform large-scale high-precision RCI calculations that would otherwise be impractical because of the overwhelming configuration-space size. It therefore holds promise for broad applications in atomic spectroscopy and tests of fundamental physics. 

\section*{Acknowledgment}
CS, ST, RS and CC acknowledge support from National Key Research and Development Project of China (No. 2022YFA1602500 and No. 2022YFA1602303) and the National Natural Science Foundation of China (No. 12674316 and No. 12393824). PJ acknowledges support from the Swedish research council under contract 2023-05367. The authors acknowledge Pavlo Bilous for helpful discussions.

\clearpage
%% References


\bibliographystyle{elsarticle-num}
\bibliography{reference}

@article{Bailey2015,
  author  = {J. E. Bailey and T. Nagayama and G. P. Loisel and G. A. Rochau and C. Blancard and J. Colgan and Ph. Cosse and G. Faussurier and C. J. Fontes and F. Gilleron and I. Golovkin and S. B. Hansen and C. A. Iglesias and D. P. Kilcrease and J. J. MacFarlane and R. C. Mancini and S. N. Nahar and C. Orban and J.-C. Pain and A. K. Pradhan and M. Sherrill and B. G. Wilson},
  title   = {A higher-than-predicted measurement of iron opacity at solar interior temperatures},
  journal = {Nature},
  volume  = {517},
  pages   = {56--59},
  year    = {2015},
  doi     = {10.1038/nature14048}
}

@article{Ullmann2017,
  author  = {J. Ullmann and Z. Andelkovic and C. Brandau and A. Dax and W. Geithner and C. Geppert and C. Gorges and M. Hammen and V. Hannen and S. Kaufmann and K. K{\"o}nig and Y. A. Litvinov and M. Lochmann and B. Maa{\ss} and J. Meisner and T. Murb{\"o}ck and R. S{\'a}nchez and M. Schmidt and S. Schmidt and M. Steck and Th. St{\"o}hlker and R. C. Thompson and C. Trageser and J. Vollbrecht and C. Weinheimer and W. N{\"o}rtersh{\"a}user},
  title   = {High precision hyperfine measurements in bismuth challenge bound-state strong-field {QED}},
  journal = {Nature Communications},
  volume  = {8},
  pages   = {15484},
  year    = {2017},
  doi     = {10.1038/ncomms15484}
}

@article{Sanner2019,
  author  = {C. Sanner and N. Huntemann and R. Lange and C. Tamm and E. Peik and M. S. Safronova and S. G. Porsev},
  title   = {Optical clock comparison for {Lorentz} symmetry testing},
  journal = {Nature},
  volume  = {567},
  pages   = {204--208},
  year    = {2019},
  doi     = {10.1038/s41586-019-0972-2}
}

@article{Godun2014,
  author  = {R. M. Godun and P. B. R. Nisbet-Jones and J. M. Jones and S. A. King and L. A. M. Johnson and H. S. Margolis and K. Szymaniec and S. N. Lea and K. Bongs and P. Gill},
  title   = {Frequency ratio of two optical clock transitions in ${}^{171}\mathrm{Yb}^{+}$ and constraints on the time variation of fundamental constants},
  journal = {Phys. Rev. Lett.},
  volume  = {113},
  pages   = {210801},
  year    = {2014},
  doi     = {10.1103/PhysRevLett.113.210801}
}

@article{King2022,
  author  = {S. A. King and L. J. Spie{\ss} and P. Micke and A. Wilzewski and T. Leopold and E. Benkler and R. Lange and N. Huntemann and A. Surzhykov and V. A. Yerokhin and J. R. {Crespo L{\'o}pez-Urrutia} and P. O. Schmidt},
  title   = {An optical atomic clock based on a highly charged ion},
  journal = {Nature},
  volume  = {611},
  pages   = {43--47},
  year    = {2022},
  doi     = {10.1038/s41586-022-05245-4}
}

@article{Zhang2024NuclearClock,
  author  = {C. Zhang and T. Ooi and J. S. Higgins and J. F. Doyle and L. von der Wense and K. Beeks and A. Leitner and G. A. Kazakov and P. Li and P. G. Thirolf and T. Schumm and J. Ye},
  title   = {Frequency ratio of the ${}^{229\mathrm{m}}\mathrm{Th}$ nuclear isomeric transition and the ${}^{87}\mathrm{Sr}$ atomic clock},
  journal = {Nature},
  volume  = {633},
  pages   = {63--70},
  year    = {2024},
  doi     = {10.1038/s41586-024-07839-6}
}

@article{FroeseFischer2019,
  author  = {C. {Froese Fischer} and G. Gaigalas and P. J{\"o}nsson and J. Biero{\'n}},
  title   = {{GRASP}2018---A {Fortran} 95 version of the general relativistic atomic structure package},
  journal = {Comput. Phys. Commun.},
  volume  = {237},
  pages   = {184--187},
  year    = {2019},
  doi     = {10.1016/j.cpc.2018.10.032}
}

@article{FroeseFischer2016,
  author  = {C. {Froese Fischer} and M. Godefroid and T. Brage and P. J{\"o}nsson and G. Gaigalas},
  title   = {Advanced multiconfiguration methods for complex atoms: {I}. {Energies} and wave functions},
  journal = {J. Phys. B: At. Mol. Opt. Phys.},
  volume  = {49},
  number  = {18},
  pages   = {182004},
  year    = {2016},
  doi     = {10.1088/0953-4075/49/18/182004}
}

@article{Jonsson2023,
  author  = {P. J{\"o}nsson and M. Godefroid and G. Gaigalas and J. Ekman and J. Grumer and W. Li and J. Li and T. Brage and I. P. Grant and J. Biero{\'n} and C. {Froese Fischer}},
  title   = {An introduction to relativistic theory as implemented in {GRASP}},
  journal = {Atoms},
  volume  = {11},
  number  = {1},
  pages   = {7},
  year    = {2023},
  doi     = {10.3390/atoms11010007}
}

@article{Si2025,
  author  = {R. Si and Y. Li and K. Wang and C. Chen and G. Gaigalas and M. Godefroid and P. J{\"o}nsson},
  title   = {{GRASPG}---{An} extension to {GRASP}2018 based on configuration state function generators},
  journal = {Comput. Phys. Commun.},
  volume  = {312},
  pages   = {109604},
  year    = {2025},
  doi     = {10.1016/j.cpc.2025.109604}
}

@article{Coe2018,
  author  = {J. P. Coe},
  title   = {Machine learning configuration interaction},
  journal = {J. Chem. Theory Comput.},
  volume  = {14},
  number  = {11},
  pages   = {5739--5749},
  year    = {2018},
  doi     = {10.1021/acs.jctc.8b00849}
}

@article{Jeong2021,
  author  = {W. Jeong and C. A. Gaggioli and L. Gagliardi},
  title   = {Active learning configuration interaction for excited-state calculations of polycyclic aromatic hydrocarbons},
  journal = {J. Chem. Theory Comput.},
  volume  = {17},
  number  = {12},
  pages   = {7518--7530},
  year    = {2021},
  doi     = {10.1021/acs.jctc.1c00769}
}

@article{Zhang2024,
  author  = {J. Zhang and J. Liu and Y. Qin and R. Liu and C. Yang and G. Jiang},
  title   = {Variational {Monte Carlo} configuration-interaction approach for multielectron atoms within the multiconfigurational {Dirac-Hartree-Fock} framework},
  journal = {Phys. Rev. A},
  volume  = {110},
  pages   = {062817},
  year    = {2024},
  doi     = {10.1103/PhysRevA.110.062817}
}

@article{Bilous2023,
  author  = {P. Bilous and A. P{\'a}lffy and F. Marquardt},
  title   = {Deep-learning approach for the atomic configuration interaction problem on large basis sets},
  journal = {Phys. Rev. Lett.},
  volume  = {131},
  pages   = {133002},
  year    = {2023},
  doi     = {10.1103/PhysRevLett.131.133002}
}

@article{Bilous2024,
  author  = {P. Bilous and C. Cheung and M. Safronova},
  title   = {Neural-network approach to running high-precision atomic computations},
  journal = {Phys. Rev. A},
  volume  = {110},
  pages   = {042818},
  year    = {2024},
  doi     = {10.1103/PhysRevA.110.042818}
}

@article{Bilous2025,
  author  = {P. Bilous and C. Cheung and M. S. Safronova},
  title   = {A neural-network-based {Python} package for performing large-scale atomic {CI} using {pCI} and other high-performance atomic codes},
  journal = {Comput. Phys. Commun.},
  volume  = {315},
  pages   = {109731},
  year    = {2025},
  doi     = {10.1016/j.cpc.2025.109731}
}

@article{Li2025,
  author  = {Yanting Li and Chaofan Shi and Ran Si and Kai Wang and Per J{\"o}nsson and Gediminas Gaigalas and Michel Godefroid and Chongyang Chen},
  title   = {Blockwise perturbative corrections in multiconfiguration calculations based on configuration-state-function generators: A revised analysis of the {W XXXVIII} spectrum},
  journal = {Phys. Rev. A},
  volume  = {111},
  pages   = {042805},
  year    = {2025},
  doi     = {10.1103/PhysRevA.111.042805}
}

@article{Li2023,
  author  = {Yan Ting Li and Kai Wang and Ran Si and Michel Godefroid and Gediminas Gaigalas and Chong Yang Chen and Per J{\"o}nsson},
  title   = {Reducing the computational load---Atomic multiconfiguration calculations based on configuration state function generators},
  journal = {Comput. Phys. Commun.},
  volume  = {283},
  pages   = {108562},
  year    = {2023},
  doi     = {10.1016/j.cpc.2022.108562}
}

@article{Gaigalas1997,
  author  = {G. Gaigalas and Z. Rudzikas and C. {Froese Fischer}},
  title   = {An efficient approach for spin-angular integrations in atomic structure calculations},
  journal = {J. Phys. B: At. Mol. Opt. Phys.},
  volume  = {30},
  number  = {17},
  pages   = {3747--3771},
  year    = {1997},
  doi     = {10.1088/0953-4075/30/17/006}
}

@article{Gustafsson2017,
  author  = {S. Gustafsson and P. J{\"o}nsson and C. {Froese Fischer} and I. P. Grant},
  title   = {Combining multiconfiguration and perturbation methods: Perturbative estimates of core-core electron correlation contributions to excitation energies in {Mg}-like iron},
  journal = {Atoms},
  volume  = {5},
  number  = {1},
  pages   = {3},
  year    = {2017},
  doi     = {10.3390/atoms5010003}
}

@article{Jonsson2014,
  author  = {P. J{\"o}nsson and P. Bengtsson and J. Ekman and S. Gustafsson and L. B. Karlsson and G. Gaigalas and C. {Froese Fischer} and D. Kato and I. Murakami and H. A. Sakaue and H. Hara and T. Watanabe and N. Nakamura and N. Yamamoto},
  title   = {Relativistic {CI} calculations of spectroscopic data for the $2p^{6}$ and $2p^{5}3l$ configurations in {Ne}-like ions between {Mg III} and {Kr XXVII}},
  journal = {At. Data Nucl. Data Tables},
  volume  = {100},
  number  = {1},
  pages   = {1--154},
  year    = {2014},
  doi     = {10.1016/j.adt.2013.06.001}
}

@article{Harris2020,
  author  = {C. R. Harris and K. J. Millman and S. J. van der Walt and R. Gommers and P. Virtanen and D. Cournapeau and E. Wieser and J. Taylor and S. Berg and N. J. Smith and R. Kern and M. Picus and S. Hoyer and M. H. van Kerkwijk and M. Brett and A. Haldane and J. {Fern{\'a}ndez del R{\'i}o} and M. Wiebe and P. Peterson and P. G{\'e}rard-Marchant and K. Sheppard and T. Reddy and W. Weckesser and H. Abbasi and C. Gohlke and T. E. Oliphant},
  title   = {Array programming with {NumPy}},
  journal = {Nature},
  volume  = {585},
  number  = {7825},
  pages   = {357--362},
  year    = {2020},
  doi     = {10.1038/s41586-020-2649-2}
}

@misc{tensorflow2015-whitepaper,
  author = {M. Abadi and A. Agarwal and P. Barham and E. Brevdo and Z. Chen and C. Citro and G. S. Corrado and A. Davis and J. Dean and M. Devin and S. Ghemawat and I. Goodfellow and A. Harp and G. Irving and M. Isard and Y. Jia and R. Jozefowicz and L. Kaiser and M. Kudlur and J. Levenberg and D. Mane and R. Monga and S. Moore and D. Murray and C. Olah and M. Schuster and J. Shlens and B. Steiner and I. Sutskever and K. Talwar and P. Tucker and V. Vanhoucke and V. Vasudevan and F. Viegas and O. Vinyals and P. Warden and M. Wattenberg and M. Wicke and Y. Yu and X. Zheng},
  title  = {{TensorFlow}: Large-scale machine learning on heterogeneous systems},
  year   = {2015},
  url    = {https://www.tensorflow.org/},
  note   = {Software available from tensorflow.org}
}

@article{Cheung2025,
  author  = {C. Cheung and S. G. Porsev and D. Filin and M. S. Safronova and M. Wehrheim and L. J. Spie{\ss} and S. Y. Chen and A. Wilzewski and J. R. {Crespo L{\'o}pez-Urrutia} and P. O. Schmidt},
  title   = {Finding the ultranarrow ${}^{3}{P}_{2}\rightarrow{}^{3}{P}_{0}$ electric quadrupole transition in {{Ni}$^{12+}$} ion for an optical clock},
  journal = {Phys. Rev. Lett.},
  volume  = {135},
  pages   = {093002},
  year    = {2025},
  doi     = {10.1103/flwf-c2m1}
}

@article{Kozlov2018,
  author  = {M. G. Kozlov and M. S. Safronova and J. R. {Crespo L{\'o}pez-Urrutia} and P. O. Schmidt},
  title   = {Highly charged ions: Optical clocks and applications in fundamental physics},
  journal = {Rev. Mod. Phys.},
  volume  = {90},
  pages   = {045005},
  year    = {2018},
  doi     = {10.1103/RevModPhys.90.045005}
}

@misc{NIST_ASD,
  author       = {A. Kramida and Yu. Ralchenko and J. Reader and {NIST ASD Team}},
  title        = {{NIST Atomic Spectra Database} (version 5.12)},
  howpublished = {National Institute of Standards and Technology, Gaithersburg, MD},
  year         = {2024},
  url          = {https://physics.nist.gov/asd},
  doi          = {10.18434/T4W30F},
  note         = {Accessed: 19 May 2026}
}






% \end{thebibliography}

\end{document}